\documentclass[prb,twocolumn,superscriptaddress,longbibliography]{revtex4-1}
\RequirePackage[mathlines]{lineno}
\usepackage{amssymb}
\usepackage{amsmath}
\usepackage{mathrsfs}
\usepackage{graphicx}
\usepackage[usenames,dvipsnames]{xcolor}
\definecolor{goodgreen}{rgb}{0.1,0.5,0}
\definecolor{goodred}{rgb}{0.7,0,0}
\usepackage[colorlinks,urlcolor=goodgreen,citecolor=blue,linkcolor=goodred]{hyperref}

\makeatletter
\newsavebox{\@brx}
\newcommand{\llangle}[1][]{\savebox{\@brx}{\(\m@th{#1\langle}\)}%
  \mathopen{\copy\@brx\kern-0.5\wd\@brx\usebox{\@brx}}}
\newcommand{\rrangle}[1][]{\savebox{\@brx}{\(\m@th{#1\rangle}\)}%
  \mathclose{\copy\@brx\kern-0.5\wd\@brx\usebox{\@brx}}}
\makeatother

\usepackage[normalem]{ulem}

\newcommand{\dcom}[1]{\textcolor{NavyBlue}{\textsf{\textbf{#1}}}}

\setpagewiselinenumbers
\modulolinenumbers[5]

\begin{document}

\preprint{\dcom{Preprint not for distribution CONFIDENTIAL, Version of \today}}
\title{Theory for constructing effective models for electrons in generic bilayer graphene}

\author{H. Minh Lam}
\author{V. Nam Do}
\email{nam.dovan@phenikaa-uni.edu.vn}    
\affiliation{Department of Basic Science, Phenikaa Institute for Advanced Study (PIAS), A1 Building, Phenikaa University, Hanoi 10000, Vietnam}

\begin{abstract}
We present and discuss in detail practical techniques in formulating effective models to describe the dynamics of low-energy electrons in generic bilayer graphene. Starting from a tight-binding model using the $p_z$ orbital of carbon atoms as a representation basis set, we reformulate it into the problem of coupling between Bloch states defined in each graphene layer. This approach allows transferring the original problem into the determination of Bloch states in two independent material layers and coupling rules of such states. We show two schemes to parameterize coupled Bloch state vectors. For the bilayer graphene configurations of small twist angle in which the long wavelength approximation is applicable, we show that an effective Hamiltonian can be written in the canonical form of a kinetic term defined by the momentum operator and a potential term defined by the position operator. The validity of effective models of different sophistication levels and their potential application in treating various physical aspects are numerically discussed.
\end{abstract}
\maketitle

\section{Introduction}
In recent years, research on engineering two-dimensional (2D) materials has been developed intensively and extensively.\cite{Sierra_2021,Liao_2019,Sangwan_2018,bao_2019} The ultimate aim is to tailor properties of material platforms to match technical requirements. The 2D materials have been now artificially created by stacking a number of atom layers together.\cite{Rhokov_2016} The bilayer graphene system is such a typical 2D material in which the cohesion between two graphene layers is governed by the van der Waals (vdW) forces. Working with these 2D vdW materials, the electronic properties can be tuned by a number of new fashions, for example, by sliding and/or twisting the two graphene layers. By these ways, it opens a possibility of engineering the electronic properties of materials called twistronics.\cite{Carr_2017} Theoretically, conventional approaches to study the electronic properties of materials are based on the calculation using the density functional theory (DFT), empirical tight-binding (TB) methods, and also effective models. The first two are the atomic-scale approaches. They are practically built on the notion of translation invariant unit cell, which allows exploiting the Bloch theorem to parameterize all electron eigen-states by a wave vector defined in a Brillouin zone. However, for a given twist angle it is not always to find a translation invariant unit cell, except for the case of commensurate alignment between the two graphene lattices. Therefore, it limits the use of the first two approaches. Besides, even in the presence of a translation invariant unit cell, the number of atoms contained in a unit cell can be very large, particularly in the case of small twist angles. Thus, the computational power required for evaluating fundamental properties of TBGs is really demanding.\cite{Uchida_2014,Lucignano_2019,Laissardiere_2012,Morell_2010} 

Working with multiple layer vdW materials one question that naturally arises is: whether can we understand the properties of the complex system if already knowing basic properties of each individual material layer? This is actually the idea for the approach of effective models to the twisted bilayer graphene (TBG) system since 2007.\cite{Santos_2007,Shallcross_2008,Bistritzer_2011,Shallcross_2013,Santos_2012,Catarina_2019}  Though the variation in the scale of the graphene lattice constant is ignored, effective models provide surprising results in describing the electronic structure of the TBG configurations of tiny twist angles in low energy range. Effective models of different sophistication levels were derived.\cite{Rhokov_2016} For example, models similar to the Dirac Hamiltonian were established on the basis of symmetry analysis of the TBG lattice.\cite{Mele_2010,Mele_2011,Gail_2011} Techniques for establishing larger dimension Hamiltonian matrix were also proposed to obtain better quantitative results.\cite{Weckbecker_2016,Tritsaris_2020} Recently, it was found efforts to include effects of lattice relaxation and lattice vibrations.\cite{Koshino_2020} Constructing effective models is a rather technical work. However, available instructions are, in general, presented succinctly and not in detail.  It thus may lead to the difficulty in understanding and/or the lack of background for the late use of such models to investigate various aspects of the electronic properties of the TBG systems, and especially, to develop for other vdW materials. Our aim in this paper is to develop a general formulation for the construction of effective models for electrons in generic bilayer systems. Methodologically, we show that effective models can be constructed by collecting an appropriate sub-set of single layer electronic states to represent the states of electrons in the complex systems. We will especially emphasize on the detail of practical techniques as deriving appropriate thumb rules. Using single-layer Bloch states as a general representation basis set we show two schemes to parameterize such coupled Bloch states that help to isolate from the total Hamiltonian matrix a block as an effective Hamiltonian. We also show that in the limit of long wavelength the procedure of constructing such an effective Hamiltonian block can be significantly simplified. It results in a useful continuum effective model that consists of a kinetic and a potential term canonically defined by the momentum and position operators, respectively. The solutions to derived effective models are presented in detail. We also discuss the validity of effective models at several approximation levels to suggest their application in studying different aspects of the bilayer systems.

The contents of the paper are arranged as follows. In Sec. \ref{Sec_II} we present the description of the atomic lattice of the bilayer graphene. In Sec. \ref{Sec_III} we start by presenting in sub-Sec. \ref{Sec_III_A} a general formulation for the Hamiltonian based on the tight-binding description for the $p_z$ atomic orbitals as the vector basis set. The translation symmetry of the graphene lattice is exploited to transform the atomic orbital basis set to the new one of Bloch states. A selection rule for the coupling between such Bloch states in the same material layer is stated in sub-Sec. \ref{Sec_III_B}. Another selection for the coupling between Bloch states in different layers is presented in sub-Sec. \ref{Sec_III_C}. We apply these selection rules to build a procedure for collecting a subset of coupled Bloch vectors that allows to establish an effective Hamiltonian model. Practical techniques are presented and discussed in detail in sub-Sec. \ref{Sec_III_D}. In sub-Sec. \ref{Sec_III_E} we show how the developed procedure is employed to derive a continuum effective model in the limit of long wavelength. The case of commensurate TBGs is detailed in sub-Sec. \ref{Sec_III_F}. To demonstrate the correctness of the calculation procedure we present in Sec. \ref{Sec_IV} a solution for Bistritzer-MacDonald model. Finally, we present the  conclusion of the paper in Sec. \ref{Sec_V}.
\section{Lattice geometry}\label{Sec_II}
Let us consider a system of two flat graphene layers stacked together with the interlayer distance $d_{GG} = 3.35$ \AA\, and relatively twisted and displaced to each other. The twist is measured by an angle $\theta$ and the displacement is measured by a vector $\boldsymbol{\tau}$. We assume that the twisted bilayer graphene system (TBG) is made via three steps: firstly, two graphene sheets are coincidentally stacked together; secondly, the two graphene layers are relatively rotated around an axis that goes perpendicularly through the graphene sheets at the center point of the atomic hexagonal ring; and thirdly, the two graphene layers are relatively displaced via the vector $\boldsymbol{\tau}$. Accordingly, when $\theta=0$ and $\boldsymbol{\tau}=0$ the TBG system corresponds to the AA-stacked configuration. By this construction, the atomic lattice of TBG has the symmetry of the $D_3$ point group if $\boldsymbol{\tau}=0$. However, one should notice the effects of the displacement and the twist that while the former destroys all point symmetries, the latter generally breaks the translation symmetry of the complex lattices. To determine the position of each carbon atom in the TBG lattice we firstly define two basis vectors of the AA-stacked configuration. For instance, we can choose: 
\begin{equation}\label{Eq1}
\mathbf{a}_1 = a\left(\frac{\sqrt{3}}{2}\hat{\mathbf{x}}+\frac{1}{2}\hat{\mathbf{y}}\right);
\mathbf{a}_2 = a\left(\frac{\sqrt{3}}{2}\hat{\mathbf{x}}-\frac{1}{2}\hat{\mathbf{y}}\right)
\end{equation}
where $a=\sqrt{3}a_{CC}$ is the lattice constant and $a_{CC} = 1.45$ \AA\, is the distance between two nearest lattice sites; $\hat{\mathbf{x}}$ and $\hat{\mathbf{y}}$ denote the unit vectors of the Cartesian coordinate frame. These two vectors define a unit cell with the area of $S_c = \|\mathbf{a}_1\times\mathbf{a}_2\| = a^2\sin(\pi/3)$. The unit cell of the AA-stacked configuration contains 4 Carbon atoms whose positions are given by:
\begin{subequations}
\begin{align}
\mathbf{d}_{A_1} &= \mathbf{OA}_1 =  -\frac{1}{2}\mathbf{d}_1,\label{Eq2a}\\
\mathbf{d}_{B_1} &= \mathbf{OB}_1 = +\frac{1}{2}\mathbf{d}_1,\label{Eq2b}\\
\mathbf{d}_{A_2} &= \mathbf{OA}_2 = -\frac{1}{2}\mathbf{d}_1+d_{GG}\hat{\mathbf{z}},\label{Eq2c}\\
\mathbf{d}_{B_2} &= \mathbf{OB}_2 = +\frac{1}{2}\mathbf{d}_1+d_{GG}\hat{\mathbf{z}},\label{Eq2d}
\end{align}
\end{subequations}
where $\mathbf{d}_1 = (\mathbf{a}_1+\mathbf{a}_2)/3$.

We also define two reciprocal lattice vectors $\mathbf{a}_1^\star,\mathbf{a}_2^\star$ associating to the vectors $\mathbf{a}_1$ and $\mathbf{a}_2$ according to the conditions $\mathbf{a}_i^\star\mathbf{a}_j=2\pi\delta_{ij}$. For $\mathbf{a}_1$ and $\mathbf{a}_2$ given by Eq. (\ref{Eq1}) we obtain:
\begin{equation}
\mathbf{a}_1^\star = \frac{4\pi}{\sqrt{3}a}\left(\frac{1}{2}\hat{\mathbf{x}}+\frac{\sqrt{3}}{2}\hat{\mathbf{y}}\right); 
\mathbf{a}_2^\star = \frac{4\pi}{\sqrt{3}a}\left(\frac{1}{2}\hat{\mathbf{x}}-\frac{\sqrt{3}}{2}\hat{\mathbf{y}}\right).\label{Eq3}
\end{equation}
These vectors define a Brillouin zone taking the shape of a hexagon with six corner points, called the $K$ points, determined by:
\begin{subequations}
\begin{align}
\mathbf{K}_1 &=\frac{2}{3}\mathbf{a}_1^\star+\frac{1}{3}\mathbf{a}_2^\star = -\mathbf{K}_4,\label{Eq4a}\\
\mathbf{K}_2 &=\frac{1}{3}\mathbf{a}_1^\star-\frac{1}{3}\mathbf{a}_2^\star = -\mathbf{K}_5,\label{Eq4b}\\
\mathbf{K}_6 &=\frac{1}{3}\mathbf{a}_1^\star+\frac{2}{3}\mathbf{a}_2^\star= -\mathbf{K}_3.\label{Eq4c}
\end{align}
\end{subequations}
These six points are classified into two equivalent classes $ [\mathbf{K}_\xi =\mathbf{K}_1] = \{\mathbf{K}_1,\mathbf{K}_3,\mathbf{K}_5\}$ with $\xi = -1$, and $[\mathbf{K}_\xi = \mathbf{K}_4] = \{\mathbf{K}_2,\mathbf{K}_4,\mathbf{K}_6\}$ with $\xi = +1$. 
The Brillouin zone has the area of $S_\text{BZ} = \|\mathbf{a}_1^{\star}\times\mathbf{a}_2^\star\| = (4\pi/\sqrt{3}a)^2\sin(2\pi/3)$.

The position of the carbon atoms on the $\ell$ graphene layer ($\ell = 1, 2$) are determined by the vectors:
\begin{align}\label{Eq5}
    \mathbf{x}_\alpha^{(\ell)} &= \mathbf{R}^{(\ell)}+\mathbf{d}_{\alpha}^{(\ell)}+(\ell-1)\boldsymbol{\tau}, 
\end{align}
where $\alpha = A, B$ and 
\begin{equation}\label{Eq6}
    \mathbf{R}^{(\ell)} = m\mathbf{a}_1^{(\ell)}+n\mathbf{a}_2^{(\ell)}; m,n\in \mathbb{Z}
\end{equation}
is the vector defining the Bravais lattice $\Gamma^{(\ell)}$ of the $\ell$ graphene layer, and $\boldsymbol{\tau}=\eta\mathbf{a}_1^{(2)}+\zeta\mathbf{a}_2^{(2)}$ with $\eta,\zeta\in [0,1)$ is the displacement vector. The two basis vectors $\mathbf{a}_{1,2}^{(\ell)}$  of $\Gamma^{(\ell)}$ are obtained by rotating appropriately the vectors $\mathbf{a}_{1,2}$:
\begin{equation}\label{Eq7}
    \mathbf{a}_{1,2}^{(\ell)} = R_z^{(\ell)}\cdot\mathbf{a}_{1,2},
\end{equation}
where $R_z^{(\ell)}$ is a matrix representing the rotation operation of an angle $(-1)^{\ell}\theta/2$ around the twist $Oz$ axis. Denote $\Gamma^{\star(\ell)}$ the reciprocal lattice associated with the Bravais lattice $\Gamma^{(\ell)}$. It is a set of vectors given by:
\begin{equation}\label{Eq8}
    \mathbf{G}^{(\ell)} = m\mathbf{a}_1^{\star(\ell)}+n\mathbf{a}_2^{\star(\ell)}; m, n\in\mathbb{Z},
\end{equation}
where $\mathbf{a}_{1,2}^{\star(\ell)}$ are the two basis vectors defined by:
\begin{equation}\label{Eq9}
    \mathbf{a}_{1,2}^{\star(\ell)} = R_z^{(\ell)}\cdot\mathbf{a}_{1,2}^{\star}.
\end{equation}
Actually, under the twist, all vectors associated with the graphene layer $\ell$ are transformed via Eqs.(\ref{Eq5}) and (\ref{Eq9}). For instance, vector $\mathbf{d}_\alpha^{(\ell)}$ in Eq. (\ref{Eq5}) is given by $\mathbf{d}_\alpha^{(\ell)} = R_z^{(\ell)}\cdot\mathbf{d}_{\alpha_\ell}$.

There is a special case of which the bilayer graphene systems is in the commensurate stacking between two graphene layers. It means that we can find two pairs of integer numbers $(m,n)$ and $(p,q)$ such that:
\begin{equation}\label{Eq10}
    m\mathbf{a}_1^{(1)}+n\mathbf{a}_2^{(1)} = p\mathbf{a}_1^{(2)}+q\mathbf{a}_2^{(2)}. 
\end{equation}
It can be shown that this requirement is satisfied not for any twist angle $\theta$, but only for the ones satisfying the conditions $p=n,q=m$ and:
\begin{equation}\label{Eq11}
    \tan\theta = \frac{(n^2-m^2)\sin(\pi/3)}{(n^2+m^2)\cos(\pi/3)+2mn}.
\end{equation}
Treating Eq. (\ref{Eq10}) rigorously may lead to the so-called Diophantine problem as presented in Ref. [\onlinecite{Shallcross_2008}]. The commensurate stacking, characterized by Eq. (\ref{Eq10}) or Eq. (\ref{Eq11}), implies that the TBG lattice has a translation symmetry, which is defined by two basis vectors:
\begin{subequations}
\begin{align}
    \mathbf{A}_1 &= m\mathbf{a}_1^{(1)}+n\mathbf{a}_2^{(1)} = n\mathbf{a}_1^{(2)}+m\mathbf{a}_2^{(2)},\label{Eq12a}\\
    \mathbf{A}_2 &= n\mathbf{a}_1^{(1)}-(m+n)\mathbf{a}_2^{(1)} = m\mathbf{a}_1^{(2)}-(m+n)\mathbf{a}_2^{(2)}.\label{Eq12b}
\end{align}
\end{subequations}
A super-cell defined by these two vectors has the area of $S_{sc} = \|\mathbf{A}_1\times\mathbf{A}_2\| = L_A^2\sin(2\pi/3)$, where $L_A = \|\mathbf{A}_1\|=\|\mathbf{A}_2\|=a\sqrt{m^2+mn+n^2}$. We denote $\Gamma_M$ the Bravais lattice of the commensurate TBG lattice. $\Gamma_M$ therefore is a set of vectors given by $\mathbf{R}=m\mathbf{A}_1+n\mathbf{A}_2$ with $m,n\in\mathbb{Z}$. The associated reciprocal lattice $\Gamma_M^\star$ is defined by two basis vectors:
\begin{subequations}
\begin{align}
    \mathbf{A}_1^\star &= \frac{a^2}{L_A^2}\left[(m+n)\mathbf{a}_1^{\star(1)}+n\mathbf{a}_2^{\star(1)}\right] \label{Eq13a}\\
    &=\frac{a^2}{L_A^2}\left[(m+n)\mathbf{a}_1^{\star(2)}+m\mathbf{a}_2^{\star(2)}\right], \label{Eq13b}\\
    \mathbf{A}_2^\star &=\frac{a^2}{L_A^2}\left[ n\mathbf{a}_1^{\star(1)}-m\mathbf{a}_2^{\star(1)}\right] \label{Eq13c}\\
    &=\frac{a^2}{L_A^2}\left[ m\mathbf{a}_1^{\star(2)}-n\mathbf{a}_2^{\star(2)}\right] \label{Eq13d}
\end{align}
\end{subequations}
The two vectors $\mathbf{A}_{1,2}^\star$ define a mini-Brillouin zone (MBZ) with the area of $S_\text{MBZ}=\|\mathbf{A}_1^\star\times\mathbf{A}_2^\star\| = (4\pi/\sqrt{3}a)^2\sin(\pi/3)/(m^2+mn+n^2) = S_\text{BZ}/(m^2+mn+n^2)$. This result shows that the area of the Brillouin zone of monolayer graphene is an integer multiple of that of the twisted bilayer lattice, $S_\text{BZ} = (m^2+mn+n^2)S_\text{MBZ}$. Thus, it suggests that a BZ can be covered by $(m^2+mn+n^2)$ pieces of MBZ. The six $K$ points of the MBZ are given by:
\begin{subequations}
\begin{align}\label{Eq14}
    \mathbf{K}_1^M &= \frac{1}{3}(\mathbf{A}_1^\star+\mathbf{A}_2^\star) = -\mathbf{K}_4^M,\\
    \mathbf{K}_2^M &= \frac{1}{3}(-\mathbf{A}_1^\star+2\mathbf{A}_2^\star) = -\mathbf{K}_5^M,\\
    \mathbf{K}_3^M &= \frac{1}{3}(-2\mathbf{A}_1^\star+\mathbf{A}_2^\star) = -\mathbf{K}_6^M.
\end{align}
\end{subequations}
For convenience, Eqs. (\ref{Eq13a}--\ref{Eq13d}) are inversely rewritten as follows:
\begin{subequations}
\begin{align}\label{Eq15}
    \mathbf{a}_1^{\star(1)} &= m\mathbf{A}_1^\star+n\mathbf{A}_2^\star,\\
    \mathbf{a}_2^{\star(1)} &= n\mathbf{A}_1^\star-(m+n)\mathbf{A}_2^\star,\\
    \mathbf{a}_1^{\star(2)} &= n\mathbf{A}_1^\star+m\mathbf{A}_2^\star,\\
    \mathbf{a}_2^{\star(2)} &= m\mathbf{A}_1^\star-(m+n)\mathbf{A}_2^\star.
\end{align}
\end{subequations}
These equations remarkably take the form similar to those in Eqs. (\ref{Eq12a}, \ref{Eq12b}). Using these expressions we can verify this  special relationship:
\begin{equation}\label{Eq16}
    \Delta\mathbf{K}=\mathbf{K}_2^{(1)}-\mathbf{K}_2^{(2)} = (m-n)(\mathbf{K}_1^M-\mathbf{K}_2^M).
\end{equation}
It means that the $\Delta\mathbf{K}$ is always a multiple of the MBZ edge.

\begin{figure}\centering
\includegraphics[clip=true,trim=1.5cm 7cm 2cm 7cm,width=\columnwidth]{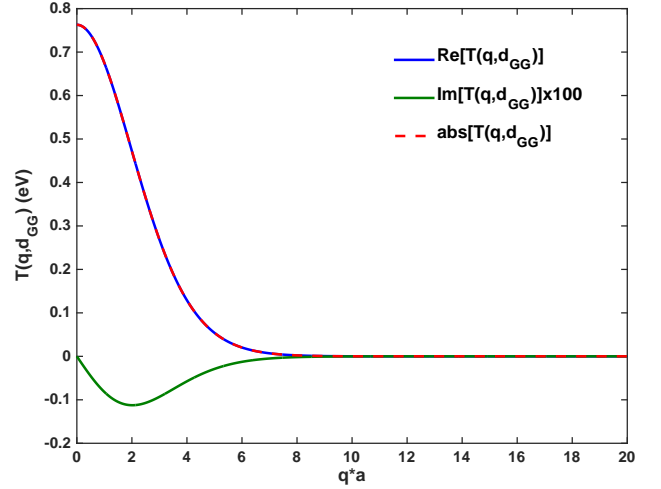}
\caption{\label{Figure_1} The Fourier image $T(\mathbf{q},d_{GG})$ of the model for the electron hopping between two graphene layers: $T(\mathbf{r}+z\mathbf{e}_z) = T(\mathbf{R}) = V_{pp\sigma}(R)(\mathbf{R}\cdot\mathbf{e}_z/R)^2+V_{pp\pi}(R)[1-(\mathbf{R}\cdot\mathbf{e}_z/R)^2]$ where  $R=\|\mathbf{R}\|$ and $V_{pp\sigma}(R) = V_{pp\sigma}^0\exp[-(R-d_{GG})/R_0]$, $V_{pp\pi}(R) = V^0_{pp\pi}\exp[-(R-a_{CC})/R_0]$. The model parameters are $V^0_{pp\pi} = -2.70$ eV, $V^0_{pp\sigma} = 0.48$ eV, $a_{CC} = 0.145$ nm, $d_{GG} = 0.335$ nm, and $R_0 = 0.148a$. Here $a = \sqrt{3}a_{CC}$ is the lattice constant of the graphene lattice.}
\end{figure}

\section{Construction of effective models}\label{Sec_III}
\subsection{Tight-binding description}\label{Sec_III_A}
In the tight-binding formalism the dynamics of an electron in an atomic lattice is described by a Hamiltonian $H$ which acts on a Hilbert space $\mathscr{H}$ spanned by a set of state vectors whose corresponding wave functions are strongly localized in the lattice. For graphene, we particularly consider a set of $p_z$-symmetry atomic orbitals $\phi_{p_z,\alpha}(\mathbf{x}-\mathbf{R}^{(\ell)}-\mathbf{d}_\alpha^{(\ell)})$ localized around the position $\mathbf{d}^{(\ell)}_\alpha$ in the graphene unit cell $\mathbf{R}^{(\ell)}$ of the material layer $\ell$. For short we denote $|\ell,\alpha,\mathbf{x}^{(\ell)}_\alpha\rangle$, where $\mathbf{x}^{(\ell)}_\alpha = \mathbf{R}^{(\ell)}+\mathbf{d}^{(\ell)}_\alpha$, the state vectors corresponding to $\phi_{p_z,\alpha}(\mathbf{x}-\mathbf{x}^{(\ell)}_\alpha)$. As the starting point we use the set of these vectors $\{|\ell,\alpha,\mathbf{R}^{(\ell)}+\mathbf{d}^{(\ell)}_\alpha\rangle\,|\, \ell = 1,2; \alpha = A, B;\mathbf{R}^{(\ell)}\in\Gamma^{(\ell)}\}$ as a representation basis, where $ \Gamma^{(\ell)}$ denotes the Bravais lattice of graphene layer $\ell$. Since each graphene layer possesses the translation symmetry, we can express $|\ell,\alpha,\mathbf{x}^{(\ell)}_\alpha\rangle$ in terms of the Fourier transform of the Bloch vectors $|\ell,\alpha,\mathbf{k}^{(\ell)}\rangle$ as follows:
\begin{equation}\label{Eq17}
    |\ell,\alpha,\mathbf{x}^{(\ell)}_\alpha\rangle=\frac{1}{\sqrt{N}}\sum_{\mathbf{k}^{(\ell)}}^{\text{BZ}_\ell}e^{i\mathbf{k}^{(\ell)}\cdot\mathbf{x}^{(\ell)}_\alpha}|\ell,\alpha,\mathbf{k}^{(\ell)}\rangle.
\end{equation}
Inversely, we have the equation:
\begin{equation}\label{Eq18}
    |\ell,\alpha,\mathbf{k}^{(\ell)}\rangle=\frac{1}{\sqrt{N}}\sum_{\mathbf{R}^{(\ell)}}^{\Gamma_\ell}e^{-i\mathbf{k}^{(\ell)}\cdot\mathbf{x}^{(\ell)}_\alpha}|\ell,\alpha,\mathbf{x}^{(\ell)}_\alpha\rangle.
\end{equation}

The single-electron Hamiltonian in the atomic lattice is, in general, determined as the summation of energies associated with all possible hopping processes of electron from a state $|\ell,\alpha,\mathbf{x}^{(\ell)}_\alpha\rangle$ to another $|\ell^\prime,\beta,\mathbf{x}^{(\ell^\prime)}_\beta\rangle$. Accordingly, we can write down the general expression for $H$ as follows:
\begin{eqnarray}\label{Eq19}
    H&=&\sum_{\ell,\ell^\prime}\sum_{\alpha,\beta}\sum_{\mathbf{R}^{(\ell)},\mathbf{R}^{(\ell^\prime)}}|\ell,\alpha,\mathbf{x}^{(\ell)}_\alpha\rangle H(\mathbf{x}^{(\ell)}_\alpha,\mathbf{x}_{\beta}^{(\ell^\prime)})\langle\ell^\prime,\beta,\mathbf{x}_{\beta}^{(\ell^\prime)}|.\nonumber\\
\end{eqnarray}
Here we denote $H(\mathbf{x}^{(\ell)}_\alpha,\mathbf{x}^{(\ell^\prime)}_{\beta}) = \langle\ell,\alpha,\mathbf{x}^{(\ell)}_\alpha|H|\ell^\prime,\beta,\mathbf{x}^{(\ell^\prime)}_\beta\rangle$ the matrix elements of the Hamiltonian in the chosen basis. These matrix elements are interpreted as the hopping integrals. All terms in Eq. (\ref{Eq19}) can be arranged into three parts: $H_1, H_2$ include the terms defining the coupling between Bloch states in the same graphene layer, and $H_{12}$ includes the terms of the coupling between Bloch states in two different graphene layers. In the tight-binding description we approximate these quantities as the two-center integrals, i.e., $H(\mathbf{x}^{(\ell)}_\alpha,\mathbf{x}^{(\ell^\prime)}_{\beta}) = T(\mathbf{x}^{(\ell)}_\alpha-\mathbf{x}^{(\ell^\prime)}_{\beta}) \equiv T(d_{\alpha\beta}^{(\ell\ell^\prime)})$ where $d_{\alpha\beta}^{(\ell\ell^\prime)} = \|\mathbf{d}_{\alpha\beta}^{(\ell\ell^\prime)}\|$ and $\mathbf{d}_{\alpha\beta}^{(\ell\ell^\prime)} = \mathbf{x}^{(\ell)}_\alpha-\mathbf{x}_{\beta}^{(\ell^\prime)}$. For numerical calculations we use the following Slater-Koster model for the hopping integrals:
\begin{equation}\label{Eq20}
    T(\mathbf{d}) = V_{pp\pi}(d)\sin^2\theta^z+V_{pp\sigma}(d)\cos^2\theta^z,
\end{equation}
where $\cos\theta^z = (\mathbf{d}\cdot\mathbf{e}_z)/d$ and
\begin{subequations}
\begin{eqnarray}
V_{pp\pi}(d) &=& V^0_{pp\pi}\exp\left(-\frac{d-a_{CC}}{r_0}\right),\label{Eq21a}\\
V_{pp\sigma}(d) &=& V^0_{pp\sigma}\exp\left(-\frac{d-d_{GG}}{r_0}\right).\label{Eq21b}
\end{eqnarray}
\end{subequations}
In this model, the parameters are commonly set to $V^0_{pp\pi} = -2.7$ eV, $V_{pp\sigma}^0 = 0.48$ eV, and $r_0 = 0.148a$.\cite{Moon-2013,Koshino-2015,Le-2018,Le-2019,Do-2019}

Using the expression given by Eq. (\ref{Eq17}), the Hamiltonian is rewritten in the form:
\begin{align}
    H = \sum_{\ell,\ell^\prime}\sum_{\alpha,\beta}|\ell,\alpha,\mathbf{k}^{(\ell)}\rangle\sum_{\mathbf{k}^{(\ell)},\mathbf{k}^{(\ell^\prime)}}^{\text{BZ}_{\ell\ell^\prime}} H_{\alpha\beta}(\mathbf{k}^{(\ell)},\mathbf{k}^{(\ell^\prime)})\langle\ell^\prime,\beta,\mathbf{k}^{(\ell^\prime)}|,\label{Eq22}
\end{align}
where we denote $H_{\alpha\beta}(\mathbf{k}^{(\ell)},\mathbf{k}^{(\ell^\prime)}) = \langle\ell,\alpha,\mathbf{k}^{(\ell)}|H|\ell^\prime,\beta,\mathbf{k}^{(\ell^\prime)}\rangle$ the matrix elements of the Hamiltonian in the basis set of Bloch vectors. The expression of $H_{\alpha\beta}(\mathbf{k}^{(\ell)},\mathbf{k}^{(\ell^\prime)})$ reads:
\begin{align}
    H_{\alpha\beta}(\mathbf{k}^{(\ell)},\mathbf{k}^{(\ell^\prime)}) =& \frac{1}{N}\sum_{\mathbf{R}^{(\ell)},\mathbf{R}^{(\ell^\prime)}}T(\mathbf{x}_\alpha^{(\ell)}-\mathbf{x}_\beta^{(\ell^\prime)})\nonumber\\
    &\hspace{1cm}\times e^{-i\mathbf{k}^{(\ell)}\cdot\mathbf{x}_\alpha^{(\ell)}}e^{i\mathbf{k}^{(\ell^\prime)}\cdot\mathbf{x}_\beta^{(\ell^\prime)}}.\label{Eq23}
\end{align}

In the following we will derive rules for the coupling between Bloch states. To do so, in the expression of $H$, we distinguish the so-called intra-layer coupling terms, $H_\ell$, including two terms when $\ell =\ell^\prime = 1$ and 2, and the so-called inter-layer coupling term, denoted by $H_{12}$, when $\ell = 1,\ell^\prime = 2$ and $\ell=2,\ell^\prime=1$.

\subsection{Rule for the intra-layer coupling}\label{Sec_III_B}
We now try to simplify the expression of $H_{\alpha\beta}(\mathbf{k}^{(\ell)},\mathbf{k}^{(\ell^\prime)})$ defined by Eq. (\ref{Eq23}). For the $H_\ell$ terms with $\ell^\prime=\ell$, we replace $\mathbf{k}^{(\ell^\prime)}$ by $\mathbf{k}^{(\ell)\prime}$. The summation over $\mathbf{x}_\alpha^{(\ell)}$ and $\mathbf{x}_\beta^{(\ell)}$ can be cast into the product of two independent sums:
\begin{align}\label{Eq24}
    H_{\alpha\beta}(\mathbf{k}^{(\ell)},\mathbf{k}^{(\ell)\prime}) &= \left(\frac{1}{N}\sum_{\mathbf{R}^{(\ell)}}e^{-i(\mathbf{k}^{(\ell)}-\mathbf{k}^{(\ell)\prime})\cdot\mathbf{x}_\alpha^{(\ell)}}\right)\nonumber\\
    &\times\left(\sum_{\mathbf{R}^{(\ell)}}T(\mathbf{x}_\alpha^{(\ell)}-\mathbf{x}_\beta^{(\ell)})e^{-i\mathbf{k}^{(\ell)\prime}\cdot(\mathbf{x}_\alpha^{(\ell)}-\mathbf{x}_\beta^{(\ell)})}\right).
\end{align}
For the first sum, by noticing Eq. (\ref{Eq5}) we generally have:
\begin{equation}\label{Eq26}
    \frac{1}{N}\sum_{\mathbf{R}^{(\ell)}}e^{-i(\mathbf{k}^{(\ell)}-\mathbf{k}^{(\ell)\prime})\cdot\mathbf{x}_\alpha^{(\ell)}} = \sum_{\mathbf{G}^{(\ell)}}\delta_{\mathbf{k}^{(\ell)}-\mathbf{k}^{(\ell)\prime},\mathbf{G}^{(\ell)}}e^{i\mathbf{G}^{(\ell)}\cdot\mathbf{d}_\alpha^{(\ell)}}.
\end{equation}
However, since both $\mathbf{k}^{(\ell)}$ and $\mathbf{k}^{(\ell)\prime}$ are restricted in the first Brillouin zone BZ$_{\ell}$, only the term with $\mathbf{G}^{(\ell)}=0$ contributes to the sum. It means that:
\begin{equation}\label{Eq27}
    \frac{1}{N}\sum_{\mathbf{R}^{(\ell)}}e^{-i(\mathbf{k}^{(\ell)}-\mathbf{k}^{(\ell)\prime})\cdot\mathbf{x}_\alpha^{(\ell)}} = \delta_{\mathbf{k}^{(\ell)},\mathbf{k}^{(\ell)\prime}}.
\end{equation}
The result implies that the dynamics of electron in each individual graphene layer is governed by only the coupling of Bloch states of the same wave vector $\mathbf{k}^{(\ell)}$. In other words, given a Bloch state with the wave vector $\mathbf{k}^{(\ell)}$, it only couples with other Bloch states of the same wave vector to form the intra-coupling Hamiltonian terms. 

For the second sum, by setting $\mathbf{d}_{\alpha\beta}^{(\ell)}=\mathbf{x}_\alpha^{(\ell)}-\mathbf{x}_\beta^{(\ell)}$, it is sufficiently interpreted as the summation over some neighborhoods of the state located at $\mathbf{x}_\alpha^{(\ell)}$. Thanks to the ``coupling rule'' just deduced, we directly calculate the second summation by counting $N_n$ neighbors at  $\{\mathbf{d}_{\alpha\beta}^{(\ell)j}\}$ of each $\mathbf{x}_\alpha^{(\ell)}$. For simplicity we set $H^{(\ell)}_{\alpha\beta}(\mathbf{k}) = H_{\alpha\beta}(\mathbf{k}^{(\ell)},\mathbf{k}^{(\ell)})$ where
\begin{eqnarray}\label{Eq25}
H^{(\ell)}_{\alpha\beta}(\mathbf{k}^{(\ell)}) &=& \sum_{j=1}^{N_n}T(\mathbf{d}^{(\ell)j}_{\alpha\beta})e^{-i\mathbf{k}^{(\ell)}\cdot\mathbf{d}_{\alpha\beta}^{(\ell)j}}.
\end{eqnarray}
Technically, the quantities $H_{\alpha\beta}^{(\ell)}(\mathbf{k}^{(\ell)})$ defined by this equation are the Fourier transforms of the hopping integrals $T(\mathbf{x})$. They comprise the elements of the Hamiltonian matrix in the representation of Bloch state vectors. Treating the coupling matrix as above and now substituting expressions (\ref{Eq27}) and (\ref{Eq25}) into Eq. (\ref{Eq22}) we obtain the expression for the intra-layer coupling Hamiltonian:
\begin{equation}\label{Eq28}
    H_{\ell} = \sum_{\mathbf{k}^{(\ell)}}^{\text{BZ}_\ell}\sum_{\alpha,\beta}|\ell,\alpha,\mathbf{k}^{(\ell)}\rangle H^{(\ell)}_{\alpha\beta}(\mathbf{k}^{(\ell)})\langle\ell,\beta,\mathbf{k}^{(\ell)}|.
\end{equation}
This Hamiltonian allows to state the following rule for the intra-coupling ($\bigstar$): Two Bloch states $|\ell,\alpha,\mathbf{k}^{(\ell)}\rangle$ and $|\ell,\beta,\mathbf{k}^{(\ell)}\rangle$ always couple to each other with the strength $H_{\alpha\beta}^{(\ell)}(\mathbf{k}^{(\ell)})$.

\subsection{Rule for the inter-layer coupling}\label{Sec_III_C}
For the inter-layer coupling terms of the Hamiltonian, the expression of $H_{\alpha\beta}(\mathbf{k}^{(\ell)},\mathbf{k}^{(\ell^\prime)})$ given in Eq. (\ref{Eq23}), in principle, can be treated in the form of Eq. (\ref{Eq24}). However, since $\mathbf{k}^{(\ell)}$ and $\mathbf{k}^{(\ell^\prime)}$ belong to different Brillouin zones, the first sum should be given by Eq. (\ref{Eq26}) rather than by Eq. (\ref{Eq27}). The expression is thus not further simplified. Importantly, in order to symmetrize the expression of $H_{\alpha\beta}(\mathbf{k}^{(\ell)},\mathbf{k}^{(\ell^\prime)})$ we should note the localization of the hopping integrals. It allows us to express $T(\mathbf{x}_\alpha^{(\ell)}-\mathbf{x}_\beta^{(\ell^\prime)})$ in the form of a Fourier series:
\begin{equation}\label{Eq29}
    T(\mathbf{x}_\alpha^{(\ell)}-\mathbf{x}_\beta^{(\ell^\prime)}) = \frac{1}{N}\sum_{\mathbf{q}\in\mathbb{R}^2}T(\mathbf{q},d_{GG})e^{i\mathbf{q}\cdot(\mathbf{x}_\alpha^{(\ell)}-\mathbf{x}_\beta^{(\ell^\prime)})},
\end{equation}
where $\mathbf{q}$ is the two-dimensional vector, $d_{GG}=z_2-z_1$, and $T(\mathbf{q},d_{GG})$ is inversely determined by
\begin{equation}\label{Eq30}
    T(\textbf{q}) = \frac{1}{S_c}\int_{S}d^2\mathbf{x}T(\mathbf{x},z)e^{-i\mathbf{q}\cdot\mathbf{x}},
\end{equation}
where $S_c = S/N$ is the area of the unit cell of the $\ell$ graphene layer. Using Eq. (\ref{Eq29}) the expression of $H_{\alpha\beta}(\mathbf{k}^{(\ell)},\mathbf{k}^{(\ell^\prime)})$ is now manipulated as follows:
\begin{widetext}
\begin{subequations}
\begin{align}
    H_{\alpha\beta}(\mathbf{k}^{(\ell)},\mathbf{k}^{(\ell^\prime)}) &= \sum_{\mathbf{q}\in\mathbb{R}^2}T(\mathbf{q},d_{GG})\frac{1}{N^2}\sum_{\mathbf{R}^{(\ell)},\mathbf{R}^{(\ell^\prime)}}e^{-i\mathbf{k}^{(\ell)}\cdot\mathbf{x}_\alpha^{(\ell)}}e^{i\mathbf{q}\cdot(\mathbf{x}_\alpha^{(\ell)}-\mathbf{x}_\beta^{(\ell^\prime)})}e^{i\mathbf{k}^{(\ell^\prime)}\cdot\mathbf{x}_\beta^{(\ell^\prime)}}\nonumber\\
    &= \sum_{\mathbf{q}\in\mathbb{R}^2}T(\mathbf{q},d_{GG})\left(\frac{1}{N}\sum_{\mathbf{R}^{(\ell)}}e^{i(\mathbf{q}-\mathbf{k}^{(\ell)})\cdot\mathbf{x}_\alpha^{(\ell)}}\right)
    \left(\frac{1}{N}\sum_{\mathbf{R}^{(\ell^\prime)}}e^{-i(\mathbf{q}-\mathbf{k}^{(\ell^\prime)})\cdot\mathbf{x}_\beta^{(\ell^\prime)}}\right)\nonumber\\
    &= \sum_{\mathbf{q}\in\mathbb{R}^2}\sum_{\mathbf{G}^{(\ell)},\mathbf{G}^{(\ell^\prime)}}T(\mathbf{q},d_{GG})e^{i\mathbf{G}^{(\ell)}\cdot\mathbf{d}_\alpha^{(\ell)}}
    e^{-i\mathbf{G}^{(\ell^\prime)}\cdot\mathbf{d}_\beta^{(\ell^\prime)}}\delta_{\mathbf{q}-\mathbf{k}^{(\ell)},\mathbf{G}^{(\ell)}}\delta_{\mathbf{q}-\mathbf{k}^{(\ell^\prime)},\mathbf{G}^{(\ell^\prime)}}\label{Eq31a}\\
    &= \sum_{\mathbf{G}^{(\ell)},\mathbf{G}^{(\ell^\prime)}}T(\mathbf{k}^{(\ell)}+\mathbf{G}^{(\ell)},d_{GG})e^{i\mathbf{G}^{(\ell)}\cdot\mathbf{d}_\alpha^{(\ell)}}e^{-i\mathbf{G}^{(\ell^\prime)}\cdot\mathbf{d}_\beta^{(\ell^\prime)}}\delta_{\mathbf{k}^{(\ell)}+\mathbf{G}^{(\ell)},\mathbf{k}^{(\ell^\prime)}+\mathbf{G}^{(\ell^\prime)}}\label{Eq31b}
\end{align}
\end{subequations}
\end{widetext}
The appearance of the Kr\"onecker symbol in Eqs. (\ref{Eq31a}) and (\ref{Eq31b}) is remarkable. It implies that not any two states $|\ell,\alpha,\mathbf{k}^{(\ell)}\rangle$ and $|\ell^\prime,\beta,\mathbf{k}^{(\ell^\prime)}\rangle$ can couple to each other, but only the ones defined by the vectors $\mathbf{k}^{(\ell)}$ and $\mathbf{k}^{(\ell^\prime)}$ related by the condition $\mathbf{k}^{(\ell)}+\mathbf{G}^{(\ell)}=\mathbf{k}^{(\ell^\prime)}+\mathbf{G}^{(\ell^\prime)}$, where  $\mathbf{G}^{(\ell)}\in\Gamma_\ell^\star$ and $\mathbf{G}^{(\ell^\prime)}\in \Gamma_{\ell^\prime}^\star$ are some vectors of the reciprocal lattices $\Gamma_\ell^\star$ and $\Gamma_{\ell^\prime}^\star$, respectively. This is the rule for the coupling between Bloch states in different graphene layers. 

\begin{figure}\centering
\includegraphics[clip=true,trim=3cm 8cm 2cm 7.5cm,width=0.85\columnwidth]{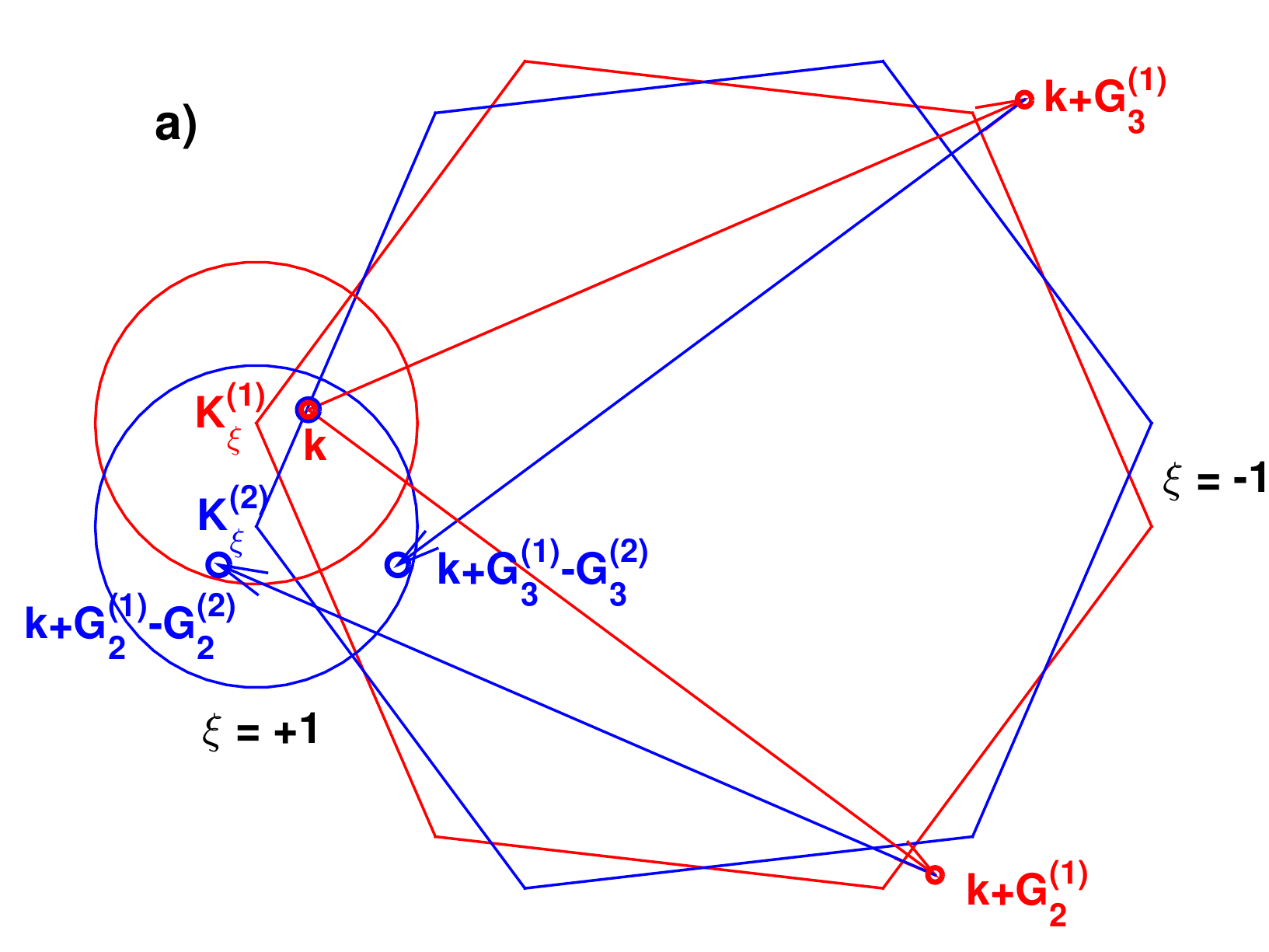}\\
\includegraphics[clip=true,trim=5cm 8.5cm 4.5cm 8cm,width=0.55\columnwidth]{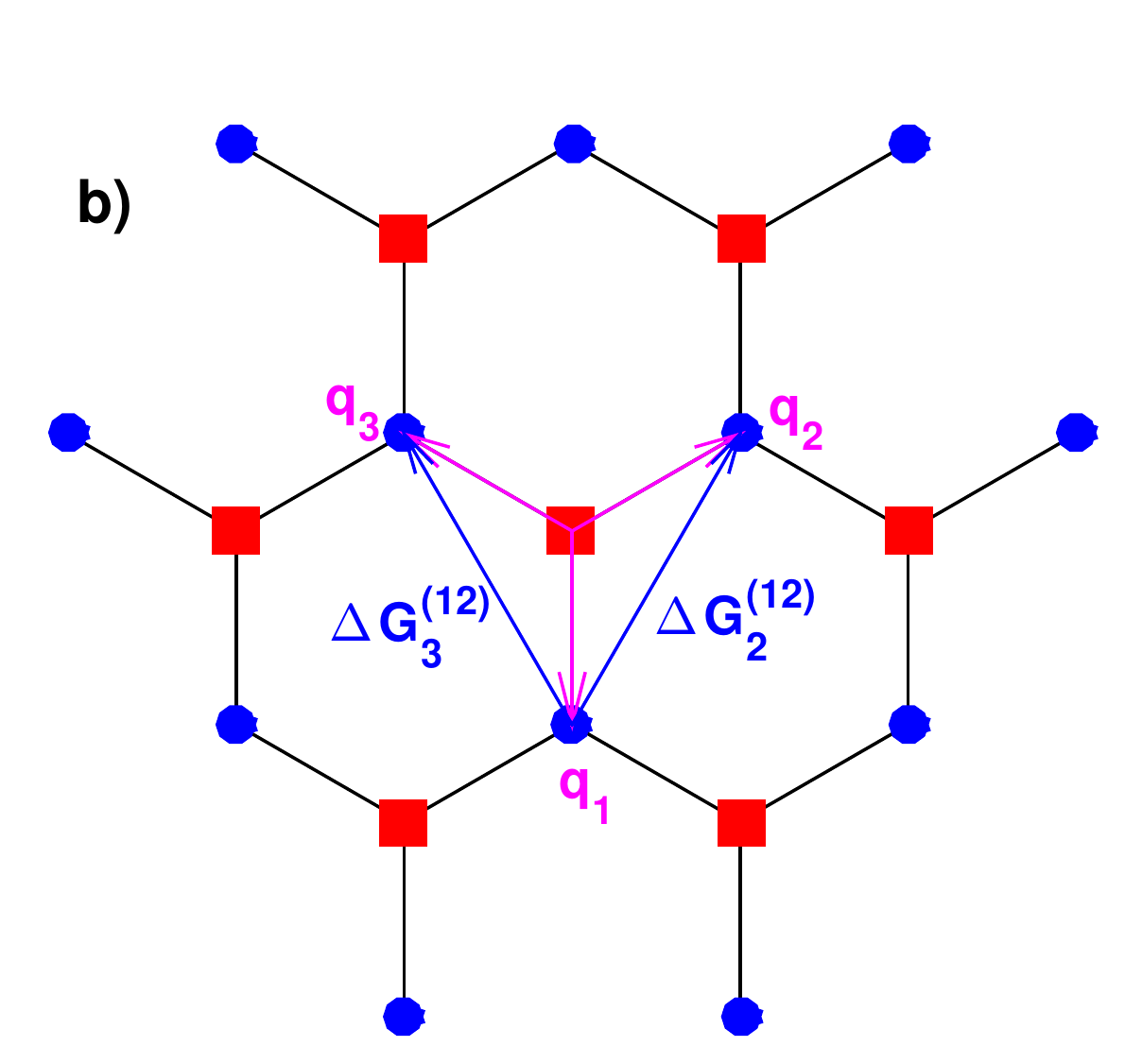}
\includegraphics[clip=true,trim=4cm 7cm 2.2cm 6.5cm,width=0.95\columnwidth]{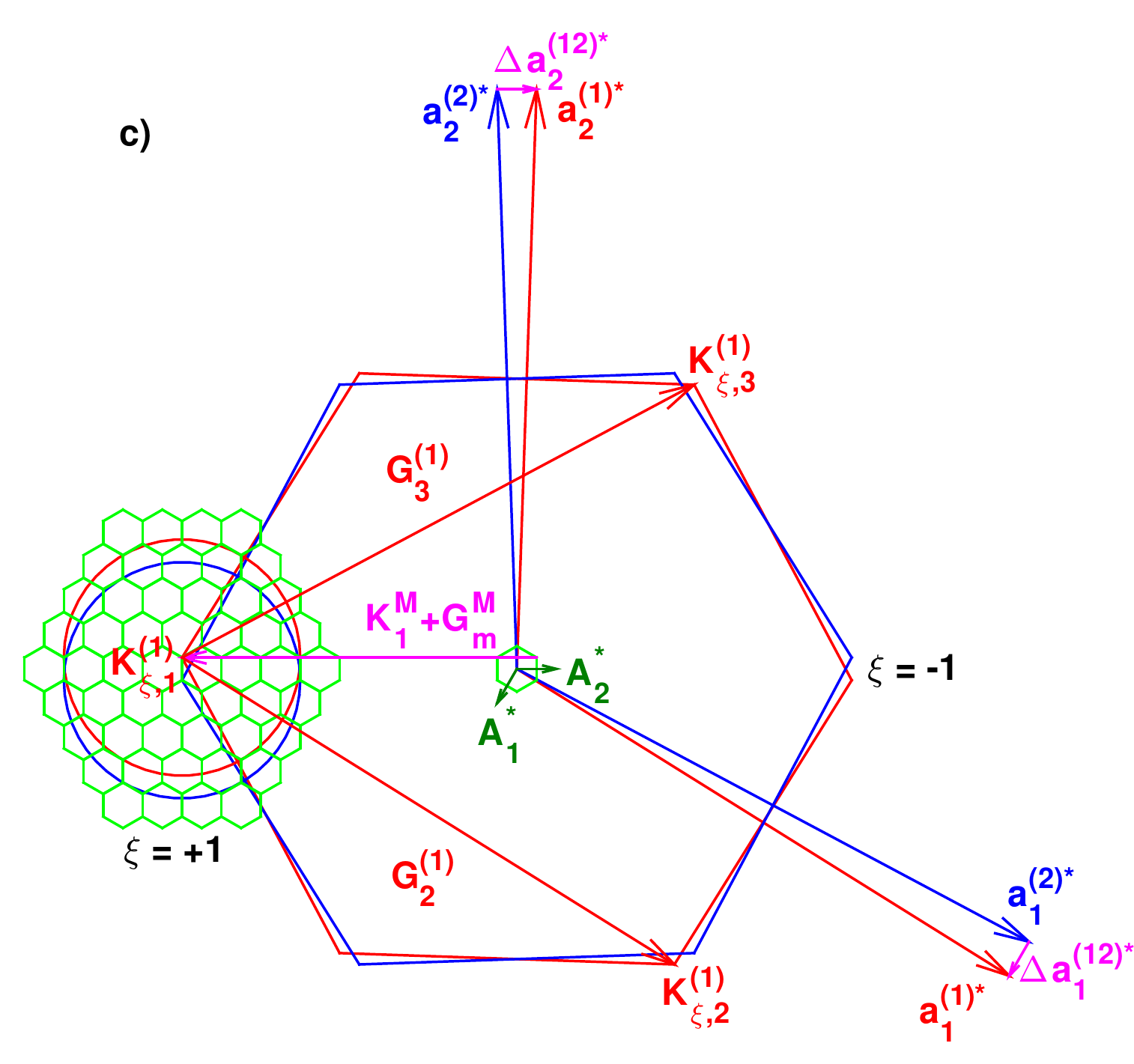}\\
\caption{\label{Figure_2} (a) Illustration for the coupling of one $\mathbf{k}^{(1)}$ point (the red dot) in the first Brillouin zone BZ$_1$ (the red hexagon) to three points $\mathbf{k}^{(2)}$ (the blue dots) in the Brillouin zone BZ$_2$ (the blue hexagon). The circles imply the domains $\mathscr{O}(\mathbf{K}^{(1)}_{\xi})$ and $\mathscr{O}(\mathbf{K}^{(2)}_{\xi})$. (b) Lattice representation for the coupling of single-layer Bloch states, see Eq. (\ref{Eq38}). The red/blue square/round nodes represent the Bloch states of the first/second graphene layer, respectively. (c) Schema for the map of the $\mathbf{q}$ points in the first Brillouin zone of the TBG lattice (the small green hexagon in the center) to the points in the $\mathscr{O}(\mathbf{K}^{(\ell)})$ domains. The map is defined by $\mathbf{q}\in \text{MBZ}\mapsto \mathbf{q}+\mathbf{G}^M_m$. The parameterization procedure using $\mathbf{q}\in\text{MBZ}$ is appropriate to the commensurate TBG configurations.}
\end{figure}

For short we denote $M_{\alpha\beta}(\mathbf{G}^{(\ell)},\mathbf{G}^{(\ell^\prime)})=e^{i\mathbf{G}^{(\ell)}\cdot\mathbf{d}_\alpha^{(\ell)}}e^{-i\mathbf{G}^{(\ell^\prime)}\cdot\mathbf{d}_\beta^{(\ell^\prime)}}$,  then substitute Eqs. (\ref{Eq31a}) and (\ref{Eq31b}) into Eq. (\ref{Eq22}). We obtain two equivalent representations for the Hamiltonian term $H_{12}$:
\begin{widetext}
\begin{align}
    H_{12} =& \sum_{\mathbf{q}\in\mathbb{R}^2}\sum_{\alpha,\beta}|1,\alpha,\underbrace{\mathbf{q}-\mathbf{G}^{(1)}}_{\mathbf{k}^{(1)}\in\text{BZ}_1}\rangle T(\mathbf{q},d_{GG})M_{\alpha\beta}({\mathbf{G}^{(1)},\mathbf{G}^{(2)}\in\text{BZ}_2})\langle 2,\beta,\underbrace{\mathbf{q}-\mathbf{G}^{(2)}}_{\mathbf{k}^{(2)}\in\text{BZ}_2}|+\nonumber\\
    &\sum_{\mathbf{q}\in\mathbb{R}^2}\sum_{\alpha,\beta}|2,\beta,\underbrace{\mathbf{q}-\mathbf{G}^{(2)}}_{\mathbf{k}^{(2)}\in\text{BZ}_2}\rangle T^\star(\mathbf{q},d_{GG})M^\star_{\alpha\beta}({\mathbf{G}^{(1)},\mathbf{G}^{(2)}})\langle 1,\alpha,\underbrace{\mathbf{q}-\mathbf{G}^{(1)}}_{\mathbf{k}^{(1)}\in\text{BZ}_1}|\label{Eq32a},\\
%
    H_{12} 
    =& \sum_{\mathbf{\mathbf{k}^{(1)}}}^{\text{BZ}_1}\sum_{\alpha,\beta}|1,\alpha,\mathbf{k}^{(1)}\rangle \left[\sum_{\mathbf{G}^{(1)}}T(\mathbf{k}^{(1)}+\mathbf{G}^{(1)})M_{\alpha\beta}(\mathbf{G}^{(1)},\mathbf{G}^{(2)})\langle 2,\beta,\underbrace{\mathbf{k}^{(1)}+\Delta\mathbf{G}^{(12)}}_{\mathbf{k}^{(2)}\in\text{BZ}_2}|\right]+\nonumber\\
    &\sum_{\mathbf{\mathbf{k}^{(2)}}}^{\text{BZ}_2}\sum_{\alpha,\beta}|2,\beta,\mathbf{k}^{(2)}\rangle \left[\sum_{\mathbf{G}^{(2)}}T(\mathbf{k}^{(2)}+\mathbf{G}^{(2)})M_{\beta\alpha}(\mathbf{G}^{(2)},\mathbf{G}^{(1)})\langle 1,\alpha,\underbrace{\mathbf{k}^{(2)}+\Delta\mathbf{G}^{(21)}}_{\mathbf{k}^{\prime(1)}\in\text{BZ}_1}|\right]\label{Eq32b},
\end{align}
\end{widetext}
wherein $\Delta\mathbf{G}^{(12)} = \mathbf{G}^{(1)}-\mathbf{G}^{(2)}$. Eq. (\ref{Eq32a}) is obtained by summing up over the vectors $\mathbf{k}^{(1)}$ and $\mathbf{k}^{(2)}$. The sum over $\mathbf{G}^{(1)}$ and $\mathbf{G}^{(2)}$ can be neglected because of the constrain that $\mathbf{q}-\mathbf{G}^{(\ell)}$ must be a point in the Brillouin zone $\text{BZ}_\ell$. In other words, $\mathbf{G}^{(1)}$ and $\mathbf{G}^{(2)}$ are two specific vectors accompanying with each value of $\mathbf{q}$ such that this condition $\mathbf{q}-\mathbf{G}^{(1,2)}\in \text{BZ}_{1,2}$ is satisfied. This equation suggests that we always find two Bloch states coupling together by starting from a point $\mathbf{q}$ in the plane $\mathbb{R}^2$ and then projecting it into the first Brillouin zone of each graphene layer to determine the corresponding vectors $\mathbf{k}^{(1)}$ and $\mathbf{k}^{(2)}$. In the same sense, the vector $\mathbf{G}^{(2)}$ in Eq. (\ref{Eq32b}) is a specific vector accompanying with each value of $\mathbf{k}^{(1)}$ and $\mathbf{G}^{(1)}$ to map the vector $\mathbf{k}^{(1)}+\mathbf{G}^{(1)}$ back into the first Brillouin zone $\text{BZ}_{2}$ of the second graphene layer, see Fig. \ref{Figure_2}(a). Eq. (\ref{Eq32b}) though does not have the symmetric form, it clearly shows a practical rule for the coupling between Bloch states in two graphene layers: for each Bloch state $|\ell,\alpha,\mathbf{k}^{(\ell)}\in\Gamma_\ell^\star\rangle$ in layer $\ell$ we always find a set of Bloch states $\{|\ell^\prime,\beta,\mathbf{k}^{(\ell)}+\Delta\mathbf{G}^{(\ell\ell^\prime)}\rangle\,|\,\forall\,\mathbf{G}^{(\ell)}\in\Gamma_\ell^\star;\mathbf{G}^{(\ell^\prime)}\in\Gamma_{\ell^\prime}^{\star}; \mathbf{k}^{(\ell)}+\Delta\mathbf{G}^{(\ell\ell^\prime)}\in\text{BZ}_{\ell^\prime}\}$ in the layer $\ell^\prime\neq\ell$ that couple to the given state ($\bigstar\bigstar$).

\subsection{Building effective models}\label{Sec_III_D}
In this sub-section we present technical points in building an effective Hamiltonian. Equation (\ref{Eq32a}) or (\ref{Eq32b}), together with Eq. (\ref{Eq28}), in principle, allow constructing a Hamiltonian matrix to describe the dynamics of electrons in its whole spectrum. Practically, we are usually interested in the behavior of electrons in a certain energy range, for example, the energy range around the Fermi energy level. It is thus expected to extract from the total Hamiltonian matrix an appropriate block whose spectrum is approximate to the exact one in the interested energy range. Technically, we need to isolate a sub-set of vectors from the basis set that dominantly couple together. The selection rule ($\bigstar\bigstar$) obviously helps to do so. Eq. (\ref{Eq32a}) shows how to recognize Bloch state vectors that couple together. They are the vectors   $|1,\alpha,\mathbf{q}-\mathbf{G}^{(1)}\rangle$ and $|2,\beta,\mathbf{q}-\mathbf{G}^{(2)}\rangle$ that are parameterized by a common vector $\mathbf{q}\in \mathbb{R}^2$. However, the matter lies in the fact that many different points $\mathbf{q}$ in the plane $\mathbb{R}^2$ can be mapped into the same point in the Brillouin zone BZ$_{1,2}$. Furthermore, though $\mathbf{q}$ varies smoothly, $\mathbf{q}-\mathbf{G}^{(1,2)}$ does not due to the discreteness of $\mathbf{G}^{(1,2)}$. Therefore, it can lead to a situation that many terms involved in Bloch states of really different energies can be arranged into the same block of the Hamiltonian matrix. It hence does not help in constructing effective models to describe the dynamics of electrons in certain narrow ranges of energy.

The second form of $H_{12}$ given by Eq. (\ref{Eq32b}) shows that for a Bloch state of one graphene layer there is a list of Bloch states of other graphene layer coupling to the given state. It is useful to see Eq. (\ref{Eq32b}) as the tight-binding form of the Hamiltonian written for Bloch states centered at discrete $\mathbf{k}$-points in the reciprocal space. It is thus impractical to consider the coupling of the state defined at the point $\mathbf{k}$ with the ones defined at the points $\mathbf{k}+\Delta\mathbf{G}^{(12)}$ far from the former. So, let us first discuss a technique to determine a set of finite number of vectors $\mathbf{G}^{(1)}$, and the corresponding $\mathbf{G}^{(2)}$, that allows approximating well the sum in Eq. (\ref{Eq32b}). In Fig. \ref{Figure_1} we show the variation of the coupling function $T(\mathbf{q},d_{GG})$ given by Eqs. (\ref{Eq20},\ref{Eq21a},\ref{Eq21b}) with respect to the length of the vector $\mathbf{q}$. Accordingly, $T(\mathbf{q},d_{GG})$ decays rapidly.  So, it is possible to define a cutoff value $G_{cutoff}$, and  $\mathbf{G}^{(1)}$ are found as the points $\mathbf{k}^{(1)}+\mathbf{G}^{(1)}$ restricted inside the circle centered at $\mathbf{k}^{(1)}$ with the radius $G_{cutoff}$. Denote $N$ the number of such $\mathbf{G}^{(1)}$ vectors. Similarly, $N$ vectors of $\mathbf{G}^{(2)}$ are also found such that $\mathbf{k}^{(1)}+\Delta\mathbf{G}^{(12)}\in$ BZ$_2$. For convenience, these $\mathbf{G}^{(\ell)}$ vectors are labelled by the subscript $j$: $\{\mathbf{G}^{(\ell)}_j\,|\, j = 1,2, \hdots,N, \|\mathbf{G}^{(\ell)}_j\| \le G_{max}\}$ with $\ell=1,2$. Since the value of $T(\mathbf{k}^{(1)}+\mathbf{G}^{(1)}_j)$ is assumed to depend only on the length of $\mathbf{k}^{(1)}+\mathbf{G}^{(1)}_j$, $N$ vectors in the set $\{\mathbf{G}^{(\ell)}_j\}$ can be further classified in classes $\mathscr{G}_1^{(1)},\mathscr{G}_2^{(1)},\mathscr{G}_3^{(1)},\hdots$ whose vectors in each class make the vectors $\mathbf{k}^{(1)}+\mathbf{G}_j^{(1)}$ having the same length. For example, if $\mathbf{k}^{(1)}$ points to a corner point $\mathbf{K}^{(1)}_\xi$ of the Brillouin zone BZ$_1$, then there are three vectors $\mathbf{G}_{1}^{(1)} = 0, \mathbf{G}_{2}^{(1)} = \xi\mathbf{a}_1^{\star(1)}$ and $\mathbf{G}_3^{(1)} = \xi(\mathbf{a}_1^{\star(1)}+\mathbf{a}_2^{\star(1)})$ that can be classified into the class $\mathscr{G}^{(1)}$ since the three vectors $\mathbf{K}^{(1)}_\xi+\mathbf{G}_j^{(1)}$ take the same length of $\|\mathbf{K}_\xi^{(1)}\| = 4\pi/3a$. Following are three classes $\mathscr{G}_j^{(1)}$ for the point $\mathbf{K}^{(1)}_\xi$. The vectors $\mathbf{G}_j^{(1)}$ are given in the format of a pair of integer numbers $(p,q)$ standing for  $\mathbf{G}_j^{(1)}=p\mathbf{a}_1^{\star(1)}+q\mathbf{a}_2^{\star(1)}$:
\begin{subequations}
\begin{align}
    \mathscr{G}_{1\xi}^{(1)} =& (0,0),\xi(1,0),\xi(1,1)\label{Eq34a}\\
    \mathscr{G}_{2\xi}^{(1)} =& \xi(0,1),\xi(0,-1),\xi(2,1)\label{Eq34b}\\
    \mathscr{G}_{3\xi}^{(1)} =& \xi(-1,0),\xi(-1,-1),\xi(1,-1),\nonumber\\
    &\xi(2,0),\xi(2,2), \xi(1,2)\label{Eq34c}
\end{align}
\end{subequations}
Similarly, the set of $N$ vectors $\mathbf{G}_j^{(2)}$ is also arranged into the classes $\mathscr{G}_j^{(2)}$ in the  correspondence to the classes $\mathscr{G}_j^{(1)}$.

We now discuss in detail techniques in the construction of an effective model that can work for a certain narrow energy range $(E_{min},E_{max})$ of the whole spectrum. Denote $\mathscr{O}(\mathbf{K}^{(\ell)})$ a domain of $\mathbf{k}^{(\ell)}$ centered at some $\mathbf{K}^{(\ell)}$ point such that there exist single-layer Bloch states of energies $E_n(\mathbf{k}^{(\ell)})$ in the range $(E_{min},E_{max})$. We expect that the combination of the state vectors $\{|1,\alpha,\mathbf{k}^{(1)}\rangle \,|\,\alpha=A,B,\mathbf{k}^{(1)}\in \mathscr{O}(\mathbf{K}^{(1)})\}$ with the ones  $\{|2,\beta,\mathbf{k}^{(2)}\rangle \,|\,\beta=A,B,\mathbf{k}^{(2)}\in \mathscr{O}(\mathbf{K}^{(2)})\}$ will form the states of the complex system in the considered energy range.  It is suggested from Eq. (\ref{Eq32b}) that the coupled Bloch state vectors, which can be parameterized by a continuous vector, are collected by a procedure as follows: (1) Starting from a state vector $|1,\alpha,\mathbf{k}^{(1)}=\mathbf{k}\rangle$ with $\mathbf{k}\in\mathscr{O}(\mathbf{K}^{(1)})$ we can find $N$ states $\{|2,\beta,\mathbf{k}^{(2)}=\mathbf{k}+\Delta\mathbf{G}^{(12)}_j\rangle, j=1,2,\hdots,N\}$ that couple to the given one. Next, (2) for each state $|2,\beta,\mathbf{k}+\Delta\mathbf{G}^{(12)}_j\in\mathscr{O}(\mathbf{K}^{(2)})\rangle$ we will find $N$ corresponding coupled states $\{|1,\alpha,\mathbf{k}^{(1)}=\mathbf{k}+\Delta\mathbf{G}^{(12)}_j-\Delta\mathbf{G}^{(12)}_i\rangle,\,i=1,2,\hdots,N\}$. Repeat this two-step procedure we will collect a desired subset of vectors from the total basis set. Since $\mathbf{G}^{(\ell)}_j$ can be expanded in terms of the basis vectors of the reciprocal lattice, $\{\mathbf{a}_1^{\star(\ell)},\mathbf{a}_2^{\star(\ell)}\}$ we have $\Delta\mathbf{G}^{(12)}_j = m_1\Delta\mathbf{a}_1^{\star(12)}+m_2\Delta\mathbf{a}_2^{\star(12)}$where $m_{1,2}\in\mathbb{Z}$. We therefore realize that the coupled Bloch state vectors related to Eq. (\ref{Eq32b}) can be written in the much more instructive form as follows:
\begin{subequations}
\begin{align}
    &|\mathbf{K}^{(1)},\alpha,\mathbf{q},\mathbf{q}^{(1)}+\Delta\mathbf{G}^{(12)}_{i}\rangle, \label{Eq35a}\\
    &|\mathbf{K}^{(2)},\beta,\mathbf{q},\mathbf{q}^{(2)}+\Delta\mathbf{G}^{(12)}_j\rangle. \label{Eq35b}
\end{align}
\end{subequations}
where $\mathbf{q}^{(1)}=0; \mathbf{q}^{(2)}=\Delta\mathbf{K}^{(12)} = \mathbf{K}^{(1)}-\mathbf{K}^{(2)}$ and $\mathbf{q}$ is a continuous vector as a small parameter measured from the $\mathbf{K}^{(\ell)}$ points. This denotation is another writing of $|\ell,\alpha,\mathbf{k}^{(\ell)}\rangle$,  where $\mathbf{k}^{(\ell)}=\mathbf{K}^{(\ell)}+\mathbf{q}+\mathbf{q}^{(\ell)}+ \Delta\mathbf{G}^{(12)}_i$, but it resembles that of localized orbitals in the real-space representation. The coupled state vectors in the collected subset, which are defined by the same parameter vector $\mathbf{q}$, can be sorted in the order:
\begin{align}\label{Eq36}
    \{&\hdots,\nonumber\\
    &|\mathbf{K}^{(1)},\alpha,\mathbf{q},\mathbf{q}^{(1)}+\Delta\mathbf{G}^{(12)}_i\rangle,\nonumber\\
    &\hdots\nonumber\\
    &|\mathbf{K}^{(1)},\alpha,\mathbf{q},\mathbf{q}^{(1)}+\Delta\mathbf{G}^{(12)}_2\rangle, \nonumber\\
    &|\mathbf{K}^{(1)},\alpha,\mathbf{q},\mathbf{q}^{(1)}+\Delta\mathbf{G}^{(12)}_1\rangle,\nonumber\\
    &|\mathbf{K}^{(2)},\beta,\mathbf{q},\mathbf{q}^{(2)}+\Delta\mathbf{G}^{(12)}_1\rangle,\nonumber\\
    &|\mathbf{K}^{(2)},\beta,\mathbf{q},\mathbf{q}^{(2)}+\Delta\mathbf{G}^{(12)}_2\rangle,\nonumber\\
    &\hdots\nonumber\\
    &|\mathbf{K}^{(2)},\beta,\mathbf{q},\mathbf{q}^{(2)}+\Delta\mathbf{G}^{(12)}_j\rangle,\nonumber\\
    &\hdots
    \}
\end{align}
By this sorting the subset of Bloch vectors can be expanded to both sides. The expansion should be terminated when the vectors $\mathbf{q}+\mathbf{q}^{(1)}+\Delta\mathbf{G}^{(12)}_i$ and $\mathbf{q}+\mathbf{q}^{(2)}+\Delta\mathbf{G}^{(12)}_j$ do no longer lie in the domains $\mathscr{O}(\mathbf{K}^{(1)})$ and $\mathscr{O}(\mathbf{K}^{(2)})$, respectively. A subset of the basis vectors determined by Eq. (\ref{Eq36}) therefore allows us to build a size-small matrix, denoted by $H[\mathbf{K}^{(1)},\mathbf{K}^{(2)}]$, which is actually a block of the total Hamiltonian matrix. The effective Hamiltonian, taking only the coupling of the Bloch states defined in the vicinity of the $\mathbf{K}^{(1)}$ and $\mathbf{K}^{(2)}$ points, is formally written as the summation of three terms $H_{eff}[\mathbf{K}^{(1)},\mathbf{K}^{(2)}]=H[\mathbf{K}^{(1)}]+H[\mathbf{K}^{(2)}]+H[\mathbf{K}^{(1)},\mathbf{K}^{(2)}]$, in which:
\begin{widetext}
\begin{subequations}
\begin{align}
    H[\mathbf{K}^{(\ell)}] &= \sum_{\mathbf{q}\in\mathscr{O}(0)}\sum_{i}\sum_{\alpha,\beta}|\mathbf{K}^{(\ell)},\alpha,\mathbf{q},\mathbf{q}^{(\ell)}+\Delta\mathbf{G}^{(12)}_i\rangle H_{\alpha\beta}^{(\ell)}(\mathbf{K}^{(\ell)}+\mathbf{q}+\mathbf{q}^{(\ell)}+\Delta\mathbf{G}^{(12)}_i)\langle \mathbf{K}^{(\ell)},\beta,\mathbf{q},\mathbf{q}^{(\ell)}+\Delta\mathbf{G}^{(12)}_i|,\label{Eq37a}\\
    H[\mathbf{K}^{(1)},\mathbf{K}^{(2)}] &= \sum_{\mathbf{q}\in\mathscr{O}(0)}\sum_{i,j}\sum_{\alpha,\beta}|\mathbf{K}^{(1)},\alpha,\mathbf{q},\mathbf{q}^{(1)}+\Delta\mathbf{G}^{(12)}_i\rangle\left[\sum_s T{(\mathbf{K}^{(1)}+\mathbf{G}^{(1)}_s)}M_{\alpha\beta}(\mathbf{G}^{(1)}_s,\mathbf{G}^{(2)}_s)\right.\nonumber\\ 
    &\hspace{6cm}\left.\times\delta_{\Delta\mathbf{G}^{(12)}_j,\Delta\mathbf{G}^{(12)}_i+\Delta\mathbf{G}^{(12)}_s}\right]\langle \mathbf{K}^{(2)},\beta,\mathbf{q},\mathbf{q}^{(2)}+\Delta\mathbf{G}^{(12)}_{j}|\nonumber+\\
    &+\sum_{\mathbf{q}\in\mathscr{O}(0)}\sum_{i,j}\sum_{\alpha,\beta}|\mathbf{K}^{(2)},\beta,\mathbf{q},\mathbf{q}^{(2)}+\Delta\mathbf{G}^{(12)}_j\rangle\left[\sum_s T{(\mathbf{K}^{(2)}+\mathbf{G}^{(2)}_s)}M_{\beta\alpha}(\mathbf{G}^{(2)}_s,\mathbf{G}^{(1)}_s)\right.\nonumber\\ 
    &\hspace{6cm}\left.\times\delta_{\Delta\mathbf{G}^{(12)}_j,\Delta\mathbf{G}^{(12)}_i+\Delta\mathbf{G}^{(12)}_s}\right]\langle \mathbf{K}^{(1)},\alpha,\mathbf{q},\mathbf{q}^{(1)}+\Delta\mathbf{G}^{(12)}_i|.\label{Eq37b}
\end{align}
\end{subequations}
\end{widetext}
This effective Hamiltonian $H_{eff}[\mathbf{K}^{(1)},\mathbf{K}^{(2)}]$ defines a Schrodinger equation whose solution should be found in terms of the coupled basis vectors given in Eqs. (\ref{Eq35a},\ref{Eq35b}) as follows:
\begin{equation}\label{Eq38b}
    |\psi\rangle = \sum_{\mathbf{q}\in\mathscr{O}(0)}\sum_{\ell,\alpha,j}C_{\ell\alpha j}(\mathbf{q})|\mathbf{K}^{(\ell)},\alpha,\mathbf{q},\mathbf{q}^{(\ell)}+\Delta\mathbf{G}^{(12)}_j\rangle,
\end{equation}
where $C_{\ell\alpha j}(\mathbf{q})$ are the linear combination coefficients needed to be determined. It is expected that the found solutions could describe correctly the electron states in an interested energy range in the whole spectrum of the bilayer systems. 

The effective model constructed via the procedure above is general for the arbitrary alignment between two graphene layers. In the case that only the lattice sliding/displacement is considered, the translation symmetry of the single graphene layer is preserved in the bilayer lattice, the Brillouin zone BZ$_1$ is identical to BZ$_2$ and defines the Brillouin zone of the bilayer lattice. In this case, the vector $\mathbf{G}^{(1)}_j \equiv \mathbf{G}^{(2)}_j$ and therefore  $\Delta\mathbf{G}^{(12)}_j = 0$. From the selection rules we see the state $|1,\alpha,\mathbf{k}^{(1)}\rangle$ and the one $|2,\beta,\mathbf{k}^{(2)}\rangle$ always couple together when  $\mathbf{k}^{(1)}=\mathbf{k}^{(2)}=\mathbf{k}\in$ BZ. The set of basis state vectors collected by the procedure above allows to full-fill the whole basis set of the Bloch state vectors of two graphene layers. The summation over $\mathbf{G}^{(1)}$ and $\mathbf{G}^{(2)}$ in Eq. (\ref{Eq32b}) therefore becomes:
\begin{equation}
    H_{\alpha\beta}(\mathbf{k}^{(1)},\mathbf{k}^{(2)})=\delta_{\mathbf{k}^{(1)},\mathbf{k}^{(2)}}\sum_{\mathbf{G}}T(\mathbf{k}+\mathbf{G})e^{i\mathbf{G}\cdot(\mathbf{d}_\alpha^{(1)}-\mathbf{d}_\beta^{(2)})}.
\end{equation}
With the notice to Eq. (\ref{Eq30}) the summation over  $\mathbf{G}$ in the above equation leads to the one identical to Eq. (\ref{Eq28}). Accordingly, in the case of sliding alignment the procedure presented here results in exactly the expression of the Hamiltonian of two graphene layers.

\subsection{Bistritzer-MacDonald model}\label{Sec_III_E}
In this subsection we consider the bilayer graphene characterized by a tiny twist angle and a small displacement vector. We are going to derive a Hamiltonian to describe effectively the behavior of electrons in an energy range around the Fermi energy level. It is well known that the low energy eigen-states of electrons in a single monolayer $\ell$ are defined by the vector $\mathbf{k}^{(\ell)}$ in the vicinity of the two independent corner points $\mathbf{K}_\xi^{(\ell)}$ ($\xi = \pm 1$) of the hexagonal Brillouin zone BZ$_\ell$. Applying the theory presented in the previous sub-section we expect to derive a desired model as the construction of the linear combination of single-layer Bloch states defined in the vicinity of the $\mathbf{K}^{(\ell)}_\xi$ points. In the limit of tiny twist angles, the two points $\mathbf{K}_\xi^{(1)}$ and $\mathbf{K}_\xi^{(2)}$ are close to each other. The two domains $\mathscr{O}(\mathbf{K}^{(1)}_\xi)$ and $\mathscr{O}(\mathbf{K}^{(2)}_\xi)$ therefore can be seen as identical. We consider only the coupling of the Bloch states defined in these $\mathbf{k}$-domains. Accordingly, we have to choose $\mathbf{G}^{(2)}_j$ close to $\mathbf{G}^{(1)}_j$ so that $\Delta\mathbf{G}^{(12)}_j$ are small vectors. Since $\mathbf{K}_\xi^{(2)}=R_z(\theta)\cdot\mathbf{K}_\xi^{(1)}$,  $\mathbf{G}^{(2)}_j$ should be chosen as $\mathbf{G}^{(2)}_j = R_z(\theta)\cdot\mathbf{G}^{(1)}_j$. For simplicity, we consider only the vectors $\mathbf{G}^{(1)}$ and $\mathbf{G}^{(2)}$ in the first class $\mathscr{G}_1^{(1)}$ and $\mathscr{G}_1^{(2)}$ to construct Eq. (\ref{Eq31b}). These vectors determine three vectors $\Delta\mathbf{G}^{(12)}_j$
\begin{subequations}
\begin{align} 
    \Delta\mathbf{G}_1^{(12)} &= 0,\label{Eq36a}\\
    \Delta\mathbf{G}_2^{(12)} &= \xi\Delta\mathbf{a}_1^{\star(12)},\label{Eq36b}\\
    \Delta\mathbf{G}_3^{(12)} &= \xi(\Delta\mathbf{a}_1^{\star(12)}+\Delta\mathbf{a}_2^{\star(12)}),\label{Eq36c}
\end{align}
\end{subequations}
that define the coupling of a Bloch state in one graphene layer to three Bloch states in the other graphene layer. 
The coupling strength is evaluated by $T(\mathbf{K}_\xi^{(1)}+\mathbf{G}_j^{(1)},d_{GG})=T(4\pi/3a,d_{GG})$. We determine the factors $M_{\alpha\beta}^j$ and arrange them into the matrix form as follows:
\begin{equation} \label{Eq37}
    M^1 = \left(\begin{array}{cc}
        1 &1  \\
        1 &1 
    \end{array}\right),
    M^2 = \left(\begin{array}{cc}
        1 &\omega  \\
        \omega^* &1 
    \end{array}\right),
    M^3 = \left(\begin{array}{cc}
        1 &\omega^*  \\
        \omega &1 
    \end{array}\right),
\end{equation}
where $\omega=e^{i2\pi/3}$. Notice that the expression of these matrices depends on the choice of the coordinate frame. By defining three vectors
\begin{equation}
    \mathbf{q}_i = \Delta\mathbf{K}^{(12)}+\Delta\mathbf{G}^{(12)}_j,
\end{equation}
the subset of basis Bloch vectors can be obtained from the list as follows:
\begin{align}\label{Eq38}
    \{&|\mathbf{K}^{(1)}_\xi,\alpha,\mathbf{q},\mathbf{Q}_1=0\rangle,\nonumber\\
    &|\mathbf{K}^{(2)}_\xi,\beta,\mathbf{q},\mathbf{Q}_2=\mathbf{Q}_1+\mathbf{q}_1\rangle,\nonumber\\
    &|\mathbf{K}^{(2)}_\xi,\beta,\mathbf{q},\mathbf{Q}_3=\mathbf{Q}_1+\mathbf{q}_2\rangle,\nonumber\\
    &|\mathbf{K}^{(2)}_\xi,\beta,\mathbf{q},\mathbf{Q}_4=\mathbf{Q}_1+\mathbf{q}_3\rangle,\nonumber\\
    &|\mathbf{K}^{(1)}_\xi,\alpha,\mathbf{q},\mathbf{Q}_5=\mathbf{Q}_2-\mathbf{q}_2\rangle,\nonumber\\
    &|\mathbf{K}^{(1)}_\xi,\alpha,\mathbf{q},\mathbf{Q}_6=\mathbf{Q}_2-\mathbf{q}_3\rangle,\nonumber\\
    &|\mathbf{K}^{(1)}_\xi,\alpha,\mathbf{q},\mathbf{Q}_7=\mathbf{Q}_3-\mathbf{q}_1\rangle,\nonumber\\ 
    &|\mathbf{K}^{(1)}_\xi,\alpha,\mathbf{q},\mathbf{Q}_8=\mathbf{Q}_3-\mathbf{q}_3\rangle,\nonumber\\    
    &|\mathbf{K}^{(1)}_\xi,\alpha,\mathbf{q},\mathbf{Q}_9=\mathbf{Q}_4-\mathbf{q}_1\rangle,\nonumber\\
    &|\mathbf{K}^{(1)}_\xi,\alpha,\mathbf{q},\mathbf{Q}_{10}=\mathbf{Q}_4-\mathbf{q}_2\rangle,\nonumber\\
    &|\mathbf{K}^{(2)}_\xi,\beta,\mathbf{q},\mathbf{Q}_{11}=\mathbf{Q}_5+\mathbf{q}_1\rangle,\nonumber\\
    &|\mathbf{K}^{(2)}_\xi,\beta,\mathbf{q},\mathbf{Q}_{12}=\mathbf{Q}_5+\mathbf{q}_3\rangle = |\mathbf{K}^{(2)}_\xi,\beta,\mathbf{q},\mathbf{Q}_{10}+\mathbf{q}_1\rangle,\nonumber\\
    &|\mathbf{K}^{(2)}_\xi,\beta,\mathbf{q},\mathbf{Q}_{13}=\mathbf{Q}_6+\mathbf{q}_1\rangle,\nonumber\\
    &|\mathbf{K}^{(2)}_\xi,\beta,\mathbf{q},\mathbf{Q}_{14}=\mathbf{Q}_{6}+\mathbf{q}_2\rangle=|\mathbf{K}^{(2)}_\xi,\beta,\mathbf{q},\mathbf{Q}_{8}+\mathbf{q}_1\rangle,\nonumber\\
    &|\mathbf{K}^{(2)}_\xi,\beta,\mathbf{q},\mathbf{Q}_{15}=\mathbf{Q}_{8}+\mathbf{q}_2\rangle,\nonumber\\
    &|\mathbf{K}^{(2)}_\xi,\beta,\mathbf{q},\mathbf{Q}_{16}=\mathbf{Q}_{7}+\mathbf{q}_2\rangle,\nonumber\\
    &|\mathbf{K}^{(2)}_\xi,\beta,\mathbf{q},\mathbf{Q}_{17}=\mathbf{Q}_7+\mathbf{q}_3\rangle=|\mathbf{K}^{(2)}_\xi,\beta,\mathbf{q},\mathbf{Q}_9+\mathbf{q}_2\rangle,\nonumber\\
    &|\mathbf{K}^{(2)}_\xi,\beta,\mathbf{q},\mathbf{Q}_{18}=\mathbf{Q}_9+\mathbf{q}_3\rangle,\nonumber\\
    &|\mathbf{K}^{(2)}_\xi,\beta,\mathbf{q},\mathbf{Q}_{19}=\mathbf{Q}_{10}+\mathbf{q}_3\rangle\nonumber\\
    &\hdots\}
\end{align}
It is remarkable to notice that three vectors $\mathbf{q}_j$ point to three $\mathbf{k}$ points that make the corners of an equivalent triangle. The vectors $\mathbf{Q}_j$ appearing in the above list therefore can be arranged into an hexagonal lattice with the vectors $\Delta\mathbf{G}^{(12)}_2$ and $\Delta\mathbf{G}^{(12)}_3$ as the two lattice vectors, see Fig. \ref{Figure_2}(b). This observation was first realized by Bistritzer and MacDonald in Ref.  [\onlinecite{Bistritzer_2011}]. This lattice representation is useful to determine all possible coupled Bloch state vectors. The subset of four first state vectors was used to show the influence of one graphene layer on the electronic structure of the other graphene layer. The dependence of the Fermi velocity on the twist angle was demonstrated analytically.\cite{Bistritzer_2011} The subset of ten first state vectors was also used to show the energy band structure of TBGs with relatively large twist angles.\cite{Catarina_2019} 

Collecting a list of coupled Bloch state vectors as given in Eq. (\ref{Eq38}) to construct a finite size matrix of effective Hamiltonian is rather technical. There is another elegantly physical approach that is based on the long wavelength approximation. Accordingly, we approximate the Bloch wave functions by the plane wave functions, i.e., $\langle\mathbf{x}|\ell,\alpha,\mathbf{k}^{(\ell)}\rangle \rightarrow e^{i\mathbf{k}^{(\ell)}\cdot\mathbf{x}}/\sqrt{S}$. In the state vector denotation, this approximation is written as the decomposition $|\ell,\alpha,\mathbf{k}^{(\ell)}\rangle\rightarrow|\ell,\alpha\rangle\otimes|\mathbf{k}^{(\ell)}\rangle$. By this approximation we can define an operator $H_{12}^\xi(\hat{\mathbf{x}})$ such that the elements of the inter-layer coupling Hamiltonian $H_{\alpha\beta}^\xi(\mathbf{k}^{(1)},\mathbf{k}^{(2)}) = \langle 1,\alpha,\mathbf{k}^{(1)}|H|2,\beta,\mathbf{k}^{(2)}\rangle$ are determined by $\langle\mathbf{k}^{(1)}|\left[\langle 1,\alpha|H_{12}^\xi(\hat{\mathbf{x}})|2,\beta\rangle\right]|\mathbf{k}^{(2)}\rangle$. Here $\hat{\mathbf{x}}$ is the position operator $\hat{\mathbf{x}}|\mathbf{x}\rangle = \mathbf{x}|\mathbf{x}\rangle$, and $\langle\mathbf{x}|\mathbf{k}^{(\ell)}\rangle = e^{i\mathbf{k}^{(\ell)}\cdot\mathbf{x}}/\sqrt{S}$. Indeed, we find that the matrix elements of this operator 
\begin{equation}\label{Eq42}
    H_{12}^\xi(\hat{\mathbf{x}}) = T(\mathbf{K}_\xi^{(1)},d_{GG})\sum_{j=1}^3M^j  e^{-i\Delta\mathbf{G}_j^{(12)}\cdot\hat{\mathbf{x}}}
\end{equation}
in the basis of vectors $\{|\ell,\alpha\rangle\otimes|\mathbf{k}^{(\ell)}\rangle\}$ are identical to the expression 
\begin{equation} \label{Eq43}
    H^\xi_{\alpha\beta}(\mathbf{k}^{(1)},\mathbf{k}^{(2)}) = T(\mathbf{K}_\xi^{(1)},d_{GG})\sum_{j=1}^3M^j_{\alpha\beta}\delta_{\mathbf{k}^{(1)},\mathbf{k}^{(2)}-\Delta\mathbf{G}^{(12)}_j},
\end{equation}
This observation for the matrix elements $H_{\alpha\beta}^\xi(\mathbf{k}^{(1)},\mathbf{k}^{(2)})$ is important, but it is not yet the whole story. The rest half must lie in a similar analysis for the intra-layer coupling matrix elements $H_{\alpha\beta}^{(\ell)}(\mathbf{k}^{(\ell)})$. 
By assuming that $\mathbf{k}^{(\ell)}$ is close to $\mathbf{K}^{(\ell)}_\xi$, we can define a small $\mathbf{q}^{(\ell)}$ such that $\mathbf{k}^{(\ell)}=\mathbf{K}^{(\ell)}_\xi+\mathbf{q}^{(\ell)}$. Now linearizing the Hamiltonian matrix elements $H^{(\ell)}_{\alpha\beta}(\mathbf{K}_\xi^{(\ell)}+\mathbf{q}^{(\ell)})$ with respect to $\mathbf{q}$ we have $H^{(\ell)}_{\alpha\beta}(\mathbf{K}_\xi^{(\ell)}+\mathbf{q}^{(\ell)})\approx H^{(\ell)}_{\alpha\beta}(\mathbf{K}_\xi^{(\ell)})+\nabla H^{(\ell)}_{\alpha\beta}(\mathbf{K}_\xi)\cdot\mathbf{q}^{(\ell)} $. This result can be also written as
\begin{align}
    &H^{(\ell)}_{\alpha\beta}(\mathbf{k}^{(\ell)}) = H^{(\ell)}_{\alpha\beta}(\mathbf{K}_\xi^{(\ell)})+\nonumber\\
    &\hspace{1cm}\langle\mathbf{k}^{(\ell)}|\frac{1}{\hbar}\nabla H^{(\ell)}_{\alpha\beta}(\mathbf{K}_\xi^{(\ell)})\cdot(\hat{\mathbf{p}}-\hbar\mathbf{K}^{(\ell)}_\xi) |\mathbf{k}^{(\ell)}\rangle,
\end{align}
where $\hat{\mathbf{p}}=-i\hbar\nabla$ is nothing rather than the momentum operator as convention. It means that, we can also define the operators:
\begin{align}\label{Eq44}
    H_\ell^\xi(\hat{\mathbf{p}}) = H^{(\ell)}(\mathbf{K}^{(\ell)}_\xi)+\frac{1}{\hbar}\nabla H^{(\ell)}(\mathbf{K}_\xi^{(\ell)})\cdot(\hat{\mathbf{p}}-\hbar\mathbf{K}_\xi^{(\ell)})
\end{align}
that work in the Hilbert space spanned by plane-wave vectors. Combining Eqs. (\ref{Eq43}) and (\ref{Eq44}) it allows us to write down an effective continuum Hamiltonian operator defined through the canonical operators $\hat{\mathbf{p}}$ and $\hat{\mathbf{x}}$ that enter separately into the kinetic and potential-like terms, i.e., $H(\hat{\mathbf{p}},\hat{\mathbf{x}}) = K(\hat{\mathbf{p}})+U(\mathbf{\hat{\mathbf{x}}})$, wherein
\begin{align}
    K(\hat{\mathbf{p}}) &= \left(\begin{array}{cc}H_1^\xi(\hat{\mathbf{p}}) &0\\ 0& H_2^\xi(\hat{\mathbf{p}})\end{array}\right),\\
    U(\hat{\mathbf{x}}) &= \left(\begin{array}{cc} 0 &H_{12}^\xi(\hat{\mathbf{x}})\\ H_{12}^{\xi\dagger}(\hat{\mathbf{x}}) &0\end{array}\right),
\end{align}
where the detailed expression of $H^\xi_{1,2}(\hat{\mathbf{p}})$ is given by Eq. (\ref{Eq50}). This continuum effective Hamiltonian allows to describe the low energy states of electrons in the bilayer graphene system as the combination of single-layer Bloch states defined in the valleys $\mathbf{K}^{(1,2)}_\xi$. This obtained Hamiltonian $H(\hat{\mathbf{p}},\hat{\mathbf{x}})$ is identical to the one driven by Bistritzer and MacDonald in Ref. [\onlinecite{Bistritzer_2011}]. It is worth noticing that the derivation of the Bistritzer-MacDonald here does not rely rigidly on the twist only, but on the genetic alignment of two graphene lattices. The layer displacement effects are, in fact, easily included in the model by modifying the elements of the matrix $M^j$ with the displacement of the atom position $\mathbf{d}_\beta^{(\ell)}\rightarrow \mathbf{d}_\beta^{(\ell)}+(\ell-1)\boldsymbol{\tau}$.

\subsection{Continuum model for commensurate TBGs}\label{Sec_III_F}
For commensurate TBG configurations, there is another way to parameterize the coupled single-layer Bloch state vectors using a vector $\mathbf{q}$ defined in the vicinity of the $\Gamma$ point rather than the $\mathbf{K}^{(\ell)}$ points. Indeed, because of the translation symmetry, the Bloch states of electron in the complex atomic lattice should be characterized by a vector $\mathbf{q}$ restricted in a zone, named mini Brillouin zone (MBZ), which is a mini-zone locating at the centre of the first Brillouin zone of the two graphene layers. The low energy states of electrons in the TBG lattice are found as the linear combination of single-layer Bloch states with the wave vector $\mathbf{k}^{(1,2)}$ defined around the point $\mathbf{K}_\xi^{(1,2)}$. We show in Fig. \ref{Figure_2} that the MBZ of the commensurate TBG lattice can be expanded to cover the BZ of two graphene layers. Especially, the corner $\mathbf{K}^M_{\xi}$ points of the MBZ can be mapped exactly to the $\mathbf{K}^{(\ell)}_\xi$ points of the BZ$_\ell$. So, in order to construct an effective Hamiltonian to determine the states of electrons that are defined by the parameter vector $\mathbf{q}\in$ MBZ we need to map each point $\mathbf{q}$ inside the MBZ to as many as possible points in the vicinity of the $\mathbf{K}^{(\ell)}_\xi$ points. This map is realized thanks to a finite set of the TBG reciprocal lattice vectors $\{\mathbf{G}^M_m\in\Gamma_M^\star\}$. By this way, from Eq. (\ref{Eq31b}) by replacing $\mathbf{k}^{(\ell)}$ by $ \mathbf{q}+\mathbf{G}^M_m$ and $\mathbf{k}^{(\ell^\prime)}$ by $\mathbf{q}^\prime+\mathbf{G}^M_n$ and the sum over $\mathbf{k}$ by the sums over $\mathbf{q}\in\text{MBZ}$ and over a set of $\{\mathbf{G}^M_m\in\Gamma^\star_M, m = 1, 2, \hdots\}$ we obtain the matrix elements of the Hamiltonian:
\begin{align}
    &H^{\ell\ell^\prime}_{\alpha\beta;mn}(\mathbf{q}) = \langle\ell,\alpha,\mathbf{q}+\mathbf{G}^M_m|H|\ell^\prime,\beta,\mathbf{q}+\mathbf{G}^M_n\rangle\nonumber\\
    &\hspace{0.2cm}=\sum_{s}T(\mathbf{K}^{(\ell)}_{\xi }+\mathbf{G}^{(\ell)}_s)M_{\alpha\beta}(\mathbf{G}^{(\ell)}_s,\mathbf{G}^{(\ell^\prime)}_s)\delta_{\mathbf{G}^M_n,\mathbf{G}^M_m-\Delta\mathbf{G}^{(\ell\ell^\prime)}_s}. \label{Eq49a}
\end{align}
The eigen-vectors of the obtained effective Hamiltonian matrix are characterized by the vector $\mathbf{q}\in\text{MBZ}$. They should be found in the form:
\begin{equation}
    |\psi(\mathbf{q})\rangle = \sum_{\ell}\sum_{\alpha}\sum_m C_{\ell,\alpha,m}(\mathbf{q})|\ell,\alpha,\mathbf{q}+\mathbf{G}^M_m\rangle.\label{Eq49b}
\end{equation}
Here we should notice that $\mathbf{q}\in$ MBZ and $\mathbf{q}+\mathbf{G}^M_m\in\mathscr{O}(\mathbf{K}_\xi^{(\ell)})$. In the long wavelength approximation, the Bloch state vectors can be approximately decomposed into  $|\ell,\alpha,\mathbf{q}+\mathbf{G}^M_m\rangle \rightarrow |\ell,\alpha\rangle\otimes|\mathbf{q}+\mathbf{G}^M_m\rangle$,  
where $|\mathbf{q}+\mathbf{G}^M_m\rangle$ are the eigen-vectors of the momentum operator $\hat{\mathbf{p}}$, and hence $|\mathbf{q}+\mathbf{G}^M_m\rangle = \exp\left(-i \mathbf{G}^{M}_m\cdot\hat{\mathbf{x}}\right)|\mathbf{q}\rangle$. Eq. (\ref{Eq49a}) is thus converted to the form of Eq. (\ref{Eq42}). Accordingly, it is worth understanding that the Bistritzer-MacDonald model is applicable to both incommensurate and commensurate TBG systems. In the next section we will present in detail the solution to the Bistritzer-MacDonald model for the commensurate TBGs. We will show how the plane-wave expansion method can be employed to solve this model.

\begin{figure}\centering
\includegraphics[clip=true,trim=1cm 7cm 2cm 7cm,width=\columnwidth]{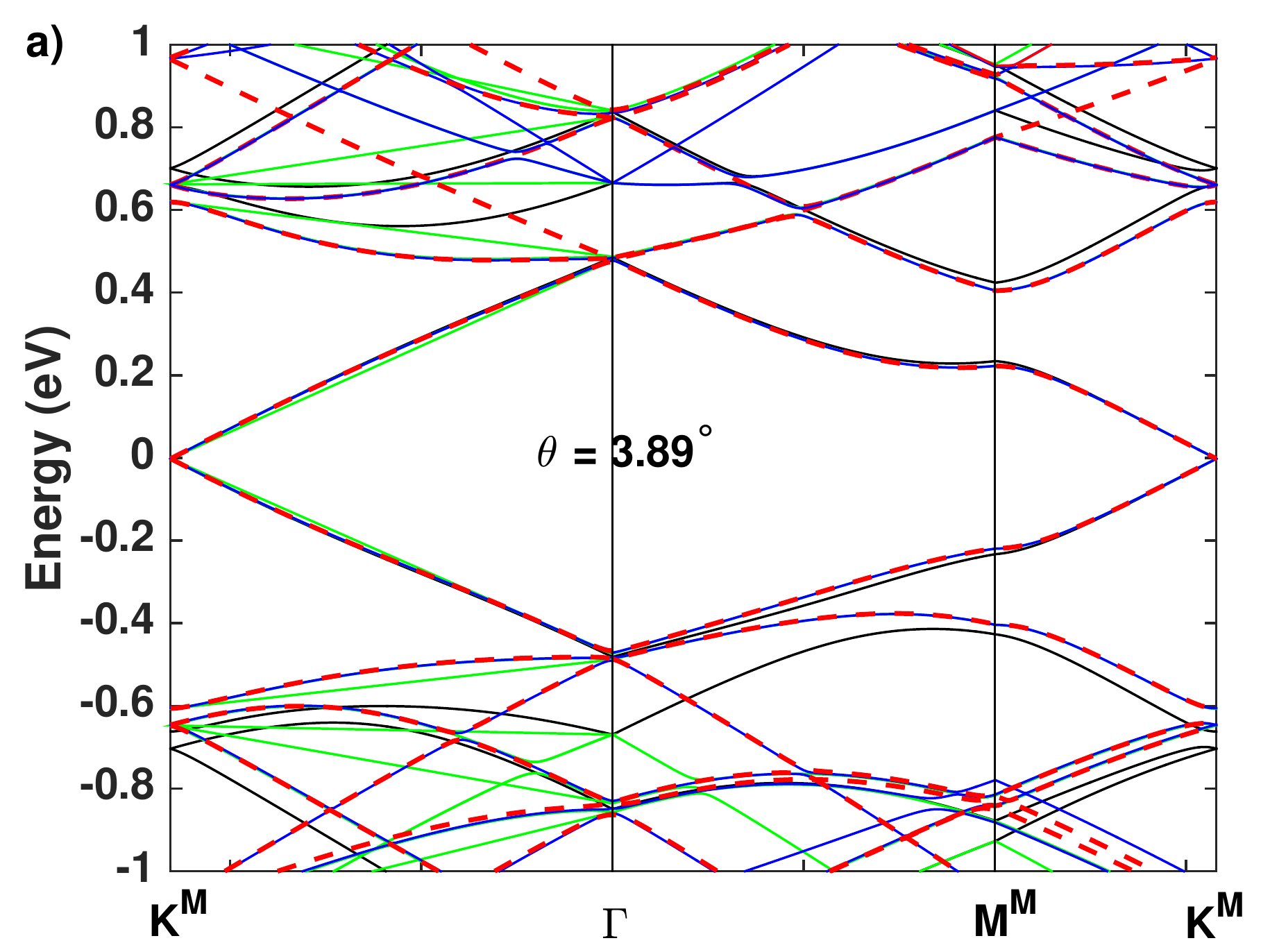}\\
\includegraphics[clip=true,trim=1cm 7cm 2cm 7cm,width=\columnwidth]{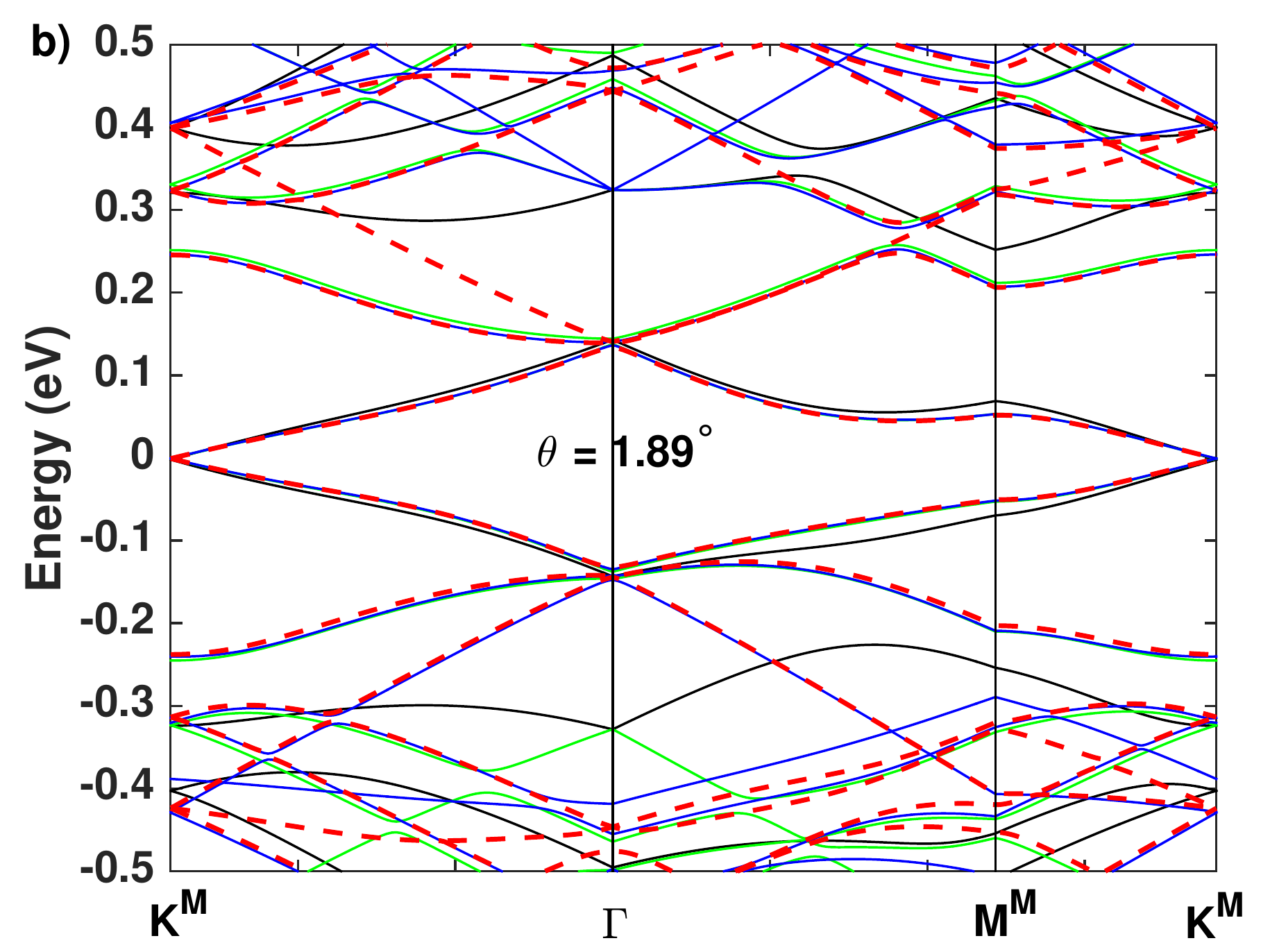}\\
\includegraphics[clip=true,trim=1cm 7cm 2cm 7cm,width=\columnwidth]{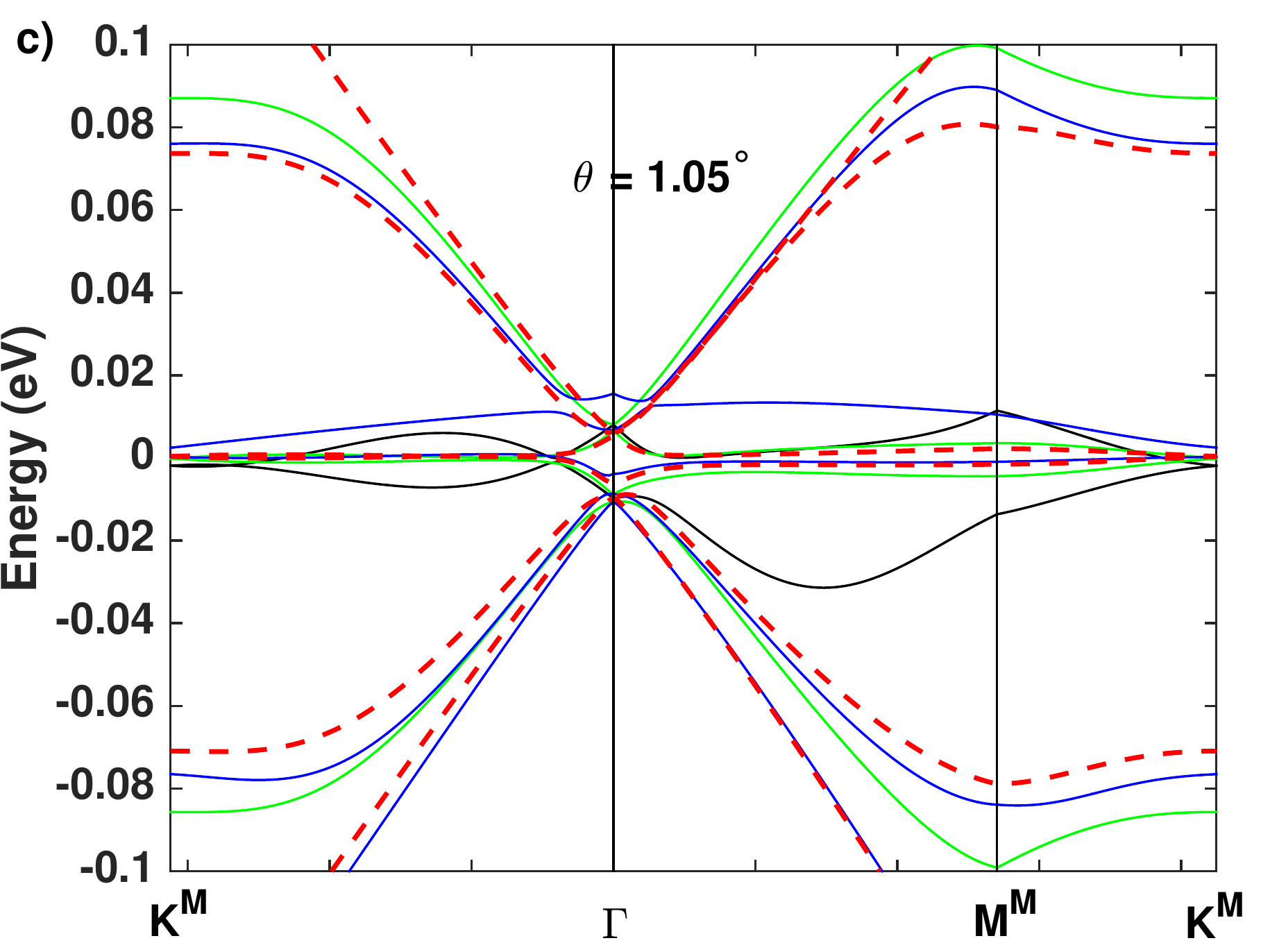}
\caption{\label{Figure_3} Energy band structures of electrons in three TBG configurations with $\theta = 3.89^\circ$ (a), $1.89^\circ$ (b), and $1.05^\circ$ (c). The black, green and blue curves are obtained using the effective models based on the number of coupled Bloch states as the basis set of $N_{\mathbf{Q}} = 4$, 10 and 19, respectively. The red curves are obtained from the model using the second scheme of parameterization with $\mathbf{q}\in $ MBZ.  The data are presented for the variation of the $\mathbf{q}$ vector in the corresponding  $\mathbf{K}^{(1)}_\xi$ valley with $\xi = +1$.}
\end{figure}

\begin{figure}\centering
\includegraphics[clip=true,trim=1.5cm 7cm 2cm 7cm,width=\columnwidth]{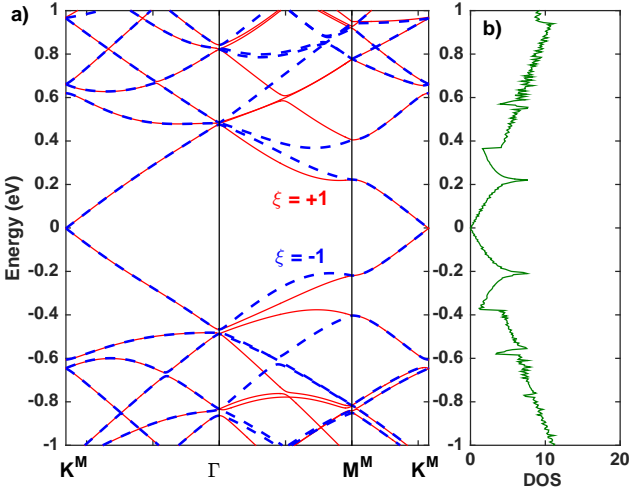}
\caption{\label{Figure_4} (a) Illustration for anisotropy of the energy band structure of the TBG system. The red-solid and blue-dashed curves are resulted for $\xi = +1$ and $\xi = -1$, respectively. (b) The total density of states.}
\end{figure}

\section{Electronic band structure}\label{Sec_IV}
\subsection{Solution of the Bistritzer-MacDonald model}
Developing effective continuum models for low-energy states of electrons in TBG systems was proceeded since 2007 by Lopes dos Santos et al. However, the model driven by Bistritzer and MacDonald in 2011 was well known and commonly used. In the previous subsections, we present the derivation of this model, and explain the key points in the two approaches. In this section, we now present our solution to the Bistritzer-MacDonald model. Accordingly, the low energy states of electrons in the TBG configurations of tiny twist angles are distinguished by the quantum indices $\xi = \pm1$ that correspond to the two nonequivalent Dirac valleys $\mathbf{K}_{\xi}^{(1,2)}$ of the graphene mono-layers.\cite{Bistritzer_2011} Notice that as the tiny twist angles, the position of the two points $\mathbf{K}_{\xi}^{(1)}$ and $\mathbf{K}_{\xi}^{(2)}$ are close to each other. The  Bistritzer-MacDonald model is given by a Hamiltonian in the real-space presentation as follows:
\begin{align}\label{Eq49}
    \hat{H}^\xi = 
    \left(\begin{array}{cc}
        H_1^\xi(\hat{\mathbf{p}}) & T^\xi(\hat{\mathbf{x}}) \\
        T^{\xi\dagger}(\hat{\mathbf{x}}) & H_2^\xi(\hat{\mathbf{p}})
    \end{array}\right)
\end{align}
Here $\hat{\mathbf{p}}$ and $\hat{\mathbf{x}}$ are the momentum operator and the position operator, respectively. The Hamiltonian $H_1^\xi(\hat{\mathbf{p}})$ and $H_2^\xi(\hat{\mathbf{p}})$ of the two uncoupled graphene layers are given by the 2D Dirac Hamiltonian:
\begin{equation}\label{Eq50}
    H_\ell^\xi(\hat{\mathbf{p}}) = - v_F(\xi\sigma_x,\sigma_y)\cdot\left[R^z\left(-\theta_\ell\right)\cdot(\hat{\mathbf{p}}-\hbar\mathbf{K}_\xi^{(\ell)})\right],
\end{equation}
where $\theta_\ell = (-1)^{\ell}\theta/2$ is the rotation angle of layer $\ell$ with respect to the fix coordinate axes $Oxy$;  $R^z\left(-\theta_\ell\right)$ is a matrix to rotate back the relevant vectors to the $Oxy$ axes to keep the canonical form of the Dirac Hamiltonian; $\mathbf{K}_\xi^{(\ell)}$ is the corner point of type (valley) $\xi$ of the first Brillouin zone of the layer $\ell$; $\sigma_x,\sigma_y$ are two conventional Pauli matrices; and $v_F$ is the Fermi velocity (The minus sign is due to the negative value of the hopping parameter, $V_{pp\pi} = -2.7$ eV). The interlayer coupling block term is given by:\cite{Koshino_2018}
\begin{align}\label{Eq51}
    T^\xi(\hat{\mathbf{x}}) =& \left(\begin{array}{cc}
    u & u^\prime \\
    u^\prime    & u 
    \end{array}\right)e^{-i\Delta\mathbf{G}^{(12)}_{\xi 1}\cdot\hat{\mathbf{x}}}+\nonumber\\
    &\left(\begin{array}{cc}
    u & u^\prime\omega^{-\xi} \\
    u^\prime\omega^{\xi}    & u 
    \end{array}\right)e^{-i\Delta\mathbf{G}^{(12)}_{\xi 2}\cdot\hat{\mathbf{x}}}+\nonumber\\
    &\left(\begin{array}{cc}
    u & u^\prime\omega^{\xi} \\
    u^\prime\omega^{-\xi}    & u 
    \end{array}\right)e^{-i\Delta\mathbf{G}^{(12)}_{\xi 3}\cdot\hat{\mathbf{x}}},
\end{align}
where $\omega = \exp(i2\pi/3)$ and $\Delta\mathbf{G}^{(12)}_{\xi 1} = 0, \Delta\mathbf{G}^{(12)}_{\xi 2} = \xi(\mathbf{a}_1^{\star(1)}-\mathbf{a}_1^{\star(2)})$ and $\Delta\mathbf{G}^{(12)}_{\xi 3} = \xi(\mathbf{a}_1^{\star(1)}-\mathbf{a}_1^{\star(2)})+\xi(\mathbf{a}_2^{\star(1)}-\mathbf{a}_2^{\star(2)})$. For commensurate TBG lattices we determine further that $\Delta\mathbf{G}^{(12)}_{\xi} = 0, \Delta\mathbf{G}^{(12)}_{\xi 2} = \mathbf{A}_1^\star$ and $\Delta\mathbf{G}^{(12)}_{\xi 3} = \xi(\mathbf{A}_1^\star+\mathbf{A}_2^\star)$, where $\mathbf{A}_{1,2}^\star$ are the basis vectors of the TBG reciprocal lattice (see Sec. I). Here we use the values proposed by Koshino for the two parameters $u = 0.0797$ eV and $u^\prime = 0.0975$ eV. The difference between these two parameters was discussed to capture effects of the lattice relaxation.\cite{Koshino_2020}

We find the spectrum of the Hamiltonian $\hat{H}$ by solving this secular equation:
\begin{align}\label{Eq52}
    \left(\begin{array}{cc}
        H_1^\xi(\hat{\mathbf{p}}) & T^\xi(\hat{\mathbf{x}}) \\
        T^{\xi\dagger}(\hat{\mathbf{x}}) & H_2^\xi(\hat{\mathbf{p}})
    \end{array}\right)|\psi_\mathbf{\xi,k}\rangle = E|\psi_{\xi,\mathbf{k}}\rangle.
\end{align}
Due to the periodicity of the moire lattice of TBG lattices, electron states are determined as the Bloch state vectors $|\psi_{\xi,\mathbf{k}}\rangle$, where $\mathbf{k}\in $ MBZ. These state vectors are expanded in terms of plane-wave vectors as follows:
\begin{equation}\label{Eq53}
    |\psi_{\xi,\mathbf{k}}\rangle = \sum_m\left(\begin{array}{c}
         C_{1,\xi,\mathbf{k}}(\mathbf{G}_m^M)  \\
         C_{2,\xi,\mathbf{k}}(\mathbf{G}_m^M)
    \end{array}\right)|\mathbf{k}+\mathbf{G}_m^M\rangle
\end{equation}
where $\hat{\mathbf{p}}|\mathbf{k}+\mathbf{G}_m^M\rangle = (\mathbf{k}+\mathbf{G}_n^M)|\mathbf{k}+\mathbf{G}_n^M\rangle$,  $\langle\mathbf{x}|\mathbf{k}+\mathbf{G}_m^M\rangle = e^{i(\mathbf{k}+\mathbf{G}_m^M)\cdot\mathbf{x}}$ and $C_{1,\xi,\mathbf{k}}(\mathbf{G}_m^M)$ and $C_{2,\xi,\mathbf{k}}(\mathbf{G}_m^M)$ are the 2D vectors of  combination coefficients that are needed to be found. Here $\{\mathbf{G}_n^M\,|\, n = 1, \hdots, N_\mathbf{G}\}$ is a set of $N_{\mathbf{G}}$ vectors of the moire reciprocal lattice. Substituting this trial expression into Eq. (\ref{Eq52}) and left-multiplying both sides with $\langle\mathbf{k}+\mathbf{G}_n^M|$ we have:
\begin{align}\label{Eq54}
    &\sum_m\langle\mathbf{k}+\mathbf{G}_n^M|\hat{H}^\xi|\mathbf{k}+\mathbf{G}_m^M\rangle\left(\begin{array}{c}
         C_{1,\xi,\mathbf{k}}(\mathbf{G}_m^M)  \\
         C_{2,\xi,\mathbf{k}}(\mathbf{G}_m^M)
    \end{array}\right) \nonumber\\
    &\hspace{3.3cm}= E\left(\begin{array}{c}
         C_{1,\xi,\mathbf{k}}(\mathbf{G}_n^M)  \\
         C_{2,\xi,\mathbf{k}}(\mathbf{G}_n^M)
    \end{array}\right)
\end{align}
With the notice that
\begin{equation}\label{Eq55}
    \langle\mathbf{k}+\mathbf{G}_n^M|H_\ell^\xi|\mathbf{k}+\mathbf{G}_m^M\rangle = H_\ell^\xi(\mathbf{k}+\mathbf{G}_n^M)\delta_{\mathbf{G}_n^M,\mathbf{G}_m^M}
\end{equation}
and
\begin{align}
    \langle\mathbf{k}+\mathbf{G}_n^M|T^\xi(\hat{\mathbf{x}})|\mathbf{k}+\mathbf{G}_m^M\rangle &= \sum_{j=1}^3T_j^\xi\delta_{\mathbf{G}_n^M,\mathbf{G}_m^M-\Delta\mathbf{G}^{(12)}_{\xi j}},\label{Eq56}\\
    \langle\mathbf{k}+\mathbf{G}_n^M|T^{\xi\dagger}(\hat{\mathbf{x}})|\mathbf{k}+\mathbf{G}_m^M\rangle &= \sum_{j=1}^3T_j^{\xi\dagger}\delta_{\mathbf{G}_n^M,\mathbf{G}_m^M+\Delta\mathbf{G}^{(12)}_{\xi j}},\label{Eq57}
\end{align}
we specify Eq. (\ref{Eq54}) in the following form:
\begin{widetext}
\begin{align}\label{Eq58}
    &\sum_m\left[\left(\begin{array}{cc}
    H_1^\xi(\mathbf{k}+\mathbf{G}_n^M)     & T_1^\xi \\
    T_1^{\xi\dagger}     & H_2^\xi(\mathbf{k}+\mathbf{G}_n^M)
    \end{array}\right)\delta_{\mathbf{G}_m^M,\mathbf{G}_n^M}+\right.\nonumber\\
    &\hspace{0.cm}+\left.\sum_{j=2}^3\left(\begin{array}{cc}
    0  & T_j^\xi \\
    0  & 0
    \end{array}\right)\delta_{\mathbf{G}_m^M,\mathbf{G}_n^M+\Delta\mathbf{G}^{(12)}_j}+\sum_{j=2}^3\left(\begin{array}{cc}
    0  & 0 \\
    T_j^{\xi\dagger}  & 0
    \end{array}\right)\delta_{\mathbf{G}_m^M,\mathbf{G}_n^M-\Delta\mathbf{G}^{(12)}_j}\right]\left(\begin{array}{c}
         C_{1,\xi,\mathbf{k}}(\mathbf{G}_m^M)  \\
         C_{2,\xi,\mathbf{k}}(\mathbf{G}_m^M)
    \end{array}\right) = E\left(\begin{array}{c}
         C_{1,\xi,\mathbf{k}}(\mathbf{G}_n^M)  \\
         C_{2,\xi,\mathbf{k}}(\mathbf{G}_n^M)
    \end{array}\right).
\end{align}
\end{widetext}
With a set of reciprocal lattice vectors $\{\mathbf{G}_n^M\,|\,n = 1,\hdots, N_\mathbf{G}\}$, the above equation is the representative of a set of $N_G$ linear equations for the coefficients $(C_{1,\xi,\mathbf{k}}(\mathbf{G}_n^M),C_{2,\xi,\mathbf{k}}(\mathbf{G}_n^M))^T$. Numerically, for each value of $\xi$ and each value of $\mathbf{k}\in \text{MBZ}$ we define a $4N_G\times 4N_G$ Hermitian matrix $H_\mathbf{k}^\xi$ that is given in the blocks as follows:
\begin{align}
    &[H_\mathbf{k}^\xi]_{n,n} = \left(\begin{array}{cc}
    H_1^\xi(\mathbf{k}+\mathbf{G}_n^M)     & T_1^\xi \\
    T_1^{\xi\dagger}     & H_2^\xi(\mathbf{k}+\mathbf{G}_n^M)
    \end{array}\right),\\
    &[H_\mathbf{k}^\xi]_{n,m_j} = \left(\begin{array}{cc}
    0     & T_j^\xi \\
    0     & 0
    \end{array}\right)\hspace{0.5cm}\text{if}\hspace{0.5cm} \mathbf{G}^M_{m_j} = \mathbf{G}^M_n+\Delta\mathbf{G}^{(12)}_{\xi j},\\
    &[H_\mathbf{k}^\xi]_{n,m_j^\prime} = \left(\begin{array}{cc}
    0     & 0 \\
    T_j^{\xi\dagger}     & 0
    \end{array}\right)\hspace{0.4cm}\text{if}\hspace{0.5cm} \mathbf{G}^M_{m_j^\prime} = \mathbf{G}^M_n-\Delta\mathbf{G}^{(12)}_{\xi j}.
\end{align}
Diagonalize the matrix $H_\mathbf{k}^\xi$ we obtain all possible eigen-values of $E_n^{\xi}(\mathbf{k})$. From these data we can display the electronic energy band structure. The issue here is the value of $N_\mathbf{G}$. Since the model is valid for low energy range in which the energy surfaces of the monolayer graphene take the cone geometry. We thus define a cutoff energy $E_c$ and then determine $N_\mathbf{G}$ the number of $\mathbf{G}^M_n$ vectors such that $\|\mathbf{G}^M_n\|\leq E_c/\hbar v_F$.

\subsection{Numerical results}
We employed the numerical method to solve Eq. (\ref{Eq58}) and similar equations. Concretely, for the parameterization of the coupled Bloch state vectors using the small vector $\mathbf{q}$ measured from th $\mathbf{K^{(\ell)}_\xi}$ points we used the approximations with the number of $\mathbf{Q}$-points  $N_\mathbf{Q}=4, 10$ and 19 to build the matrix for the effective Hamiltonian. The purpose of this is to evaluate the ability of describing electronic states of the TBG lattices from the simple to complicated levels of effective models. For commensurate TBG configurations we used the second parameterization scheme using the vector $\mathbf{q}$ defined in the mini Brillouin zone of the reciprocal lattice. Since the number of basis state vectors can be easily extended in the latter scheme, it allows to describe precisely all the energy dispersion curves in the energy range containing the Fermi energy level, i.e., ranging from -3 eV up to 3 eV. We hence use this scheme as a benchmark to validate effective models using the first scheme of parameterization of coupled Bloch states. 

When building Hamiltonian matrices we distinguish the contribution of the Bloch states defined in the nonequivalent $\mathbf{K}$ valleys with $\xi = \pm 1$ as denoted in Fig. \ref{Figure_2}. Technically, for the first scheme of parameterization the Hamiltonian matrices have the size of $2N_\mathbf{Q}\times 2N_\mathbf{Q}$ with $N_\mathbf{Q}=4$, 10 and 19, corresponding to three simple truncations of the list of coupled Bloch vectors. Meanwhile, it is $4N_\mathbf{G^M}\times 4N_\mathbf{G^M}$ for the second scheme of parameterization, where $N_\mathbf{G^M}$ is the number of the reciprocal lattice vectors of the TBG lattice to map the mini Brillouin zone MBZ to the domains $\mathscr{O}(\mathbf{K}^{(\ell)})$, see Fig. \ref{Figure_2}c. The Hamiltonian matrices are Hermitian and depend on the parameter vector $\mathbf{q}$. They are numerically diagonalized for each value of $\mathbf{q}$ to display the energy band structure. We investigated the band structure of the TBG configurations with the twist angle varying in a large range, from $1.05^\circ$ to $30^\circ$, but show in Fig. \ref{Figure_3} data for three configurations with $\theta = 3.89^\circ$ ($m=9$), $1.89^\circ$ ($m=18$) and $1.05^\circ$ ($m=32$). Our obtained results are in agreement with other available data in literature.\cite{Morell_2010,Laissardiere_2012,Moon-2013,Koshino_2018}

We realize that, for the TBG configurations with the twist angles not too small, $\theta>2.5^\circ$, the model with $N_{\mathbf{Q}}=4$ describes well two energy bands closest to the Fermi level, see the black curves in Fig. \ref{Figure_3}. However, it does not determine fully the number of bands in the energy range of $(-1,1)$ eV because there are only 8 Bloch state vectors used to represent the electronic states of the complex system. By extending $N_\mathbf{Q}$ to 10, the obtained model allows to produce correctly the number of energy bands in a narrow energy range about the Fermi level, see the green curves in Fig. \ref{Figure_3}. However, there are green curves that are not identical to the red curves, which are obtained by using the second parameterization scheme. Extending $N_\mathbf{Q}$ to 19 allows to improve quantitatively the blue energy dispersion curves to coincide with the red curves. However, deviations are still observed for the configurations with small twist angles, see Figs. 3(a) and 3(b), in the higher energy ranges, for instance, $E > 0.6$ eV and $E< -0.6$ eV for $\theta = 3.89^\circ$, and $E> 0.3$ eV and $E< -0.3$ eV for $\theta = 1.89^\circ$. For the special TBG configuration with $\theta = 1.05^\circ$ we clearly realize the failure of the effective model with $N_\mathbf{Q}=4$ in describing the “flat bands” about the Fermi level. It is surprised that the model with $N_\mathbf{Q}=10$ reproduces well these special bands, but the model with $N_\mathbf{Q}=19$ as expected does not. Notice that in the two models with $N_\mathbf{Q}=4$ and 19 there is an unbalance between the number of Bloch states in two graphene layers contributing to the electronic states of the complex system, but it is not the case in the model with $N_\mathbf{Q}=10$. It therefore suggests that in order to describe well the electronic structure of the TBG system, the effective models should be constructed on the basis set of state vectors with the balance of the number of states in each graphene layer contributing to the real system.

We sum up our investigation as follows: The model with $N_\mathbf{Q}=4$ is simple and easy to be constructed, but it allows to describe rather well the two energy bands closest to the Fermi level. This model was actually used to show analytically the dependence of the Fermi velocity on the twist angle.\cite{Santos_2007,Bistritzer_2011} Because the $8\times 8$ matrix can be arranged into a $4\times 4$ block matrix whose diagonal blocks can be approximated by the 2D Dirac model, this model can be transformed into the real-space representation. It is therefore suitable for the investigation of properties involving spatial effects of the TBG system. Extending the number of coupled Bloch states to represent the electron states in the complex systems in general allows to improve quantitatively the electronic band structure in a large energy range about the Fermi level. The models with $N_\mathbf{Q} = 10$ and $19$ involve in small size matrices (of 20 and 38 dimensions, respectively), so they are really efficient in the numerical calculation viewpoint. These models are useful to describe the electronic structure and optical properties of incommensurate TBG configurations. For commensurate TBG configurations, the second scheme of parameterization is really useful. It leads to a compact effective model that can be formally separated into a kinetic and a potential part. The so-called Bistritzer-MacDonald Hamiltonian takes this beauty, validating in the long wavelength approximation. The model is thus suitable to describe efficiently not only optical transition processes but also transport properties of electrons taking place in the bilayer lattices.

\section{Conclusion}\label{Sec_V}
We present and discuss in detail practical techniques in building effective models to describe the dynamics of electrons in certain energy ranges of the whole electronic spectrum of the generic bilayer graphene lattices. The electronic states of the bilayer system are determined as the linear combinations of single-layer Bloch states. We symmetrize the expression of the interlayer coupling Hamiltonian to figure out a selection rule that allows to determine coupled Bloch states. Concretely, two Bloch states can couple to each other if the difference between two wave vectors must be equal to the difference of two reciprocal lattice vectors of the two layers. On the basis of this selection rule, we present a procedure to  collect a subset of coupled Bloch states. This subset of states allows to isolate a block from the total Hamiltonian matrix to describe the dynamics of electrons in a certain narrow energy range. In the approximation of long wavelength, we show that when the Bloch functions in the collected subset are replaced by the plane wave functions, an effective continuum model can be established. Interestingly, this model is explicitly defined through the momentum operator and the position operator, which do not mix together, but enter into two independent terms. That allows to write the effective Hamiltonian as the sum of the kinetic- and potential-like terms. Applying the established procedure to the twisted bilayer graphene of tiny twist angles, we recover the model established by Bistritzer-MacDonald in 2011 by considering only the strongest coupling of Bloch states defined in the $\mathbf{K}$-valleys of two graphene layers. We present in detail a plane wave expansion solution to this model and numerical results for some commensurate TBG configurations. The obtained data for the electronic structure are in good agreement with those resulted from the exhausted tight-binding and DFT calculations. The practical rules and technical discussions presented in this work are expected to provide useful methodological knowledge as the background to exploit the effective continuum Bistritzer-MacDonald model for studying various physical aspects of the bilayer graphene system as well as to develop for other multiple layer van der Waals material systems.

\bibliographystyle{apsrev4-1}
\bibliography{bibliography}

\end{document}